\documentclass{acmtrans2m}
\usepackage{latexsym}
\usepackage{amsmath}
\usepackage{amsfonts}
\usepackage{amssymb}
\usepackage{amscd}
\newcommand{\commentout}[1]{}
\newcommand{\REMARQUE}[1]{\textbf{[[#1]]}}

\newtheorem{theorem}{Theorem}[section]

\newtheorem{corollary}[theorem]{Corollary}
\newtheorem{proposition}[theorem]{Proposition}
\newtheorem{lemma}[theorem]{Lemma}
\newdef{definition}[theorem]{Definition}
\newdef{remark}[theorem]{Remark}

\newcommand{\thm}{\begin{theorem}}
\newcommand{\lem}{\begin{lemma}}
\newcommand{\pro}{\begin{proposition}}
\newcommand{\dfn}{\begin{definition}}
\newcommand{\rem}{\begin{remark}}
\newcommand{\xam}{\begin{example}}
\newcommand{\cor}{\begin{corollary}}
\newcommand{\prf}{\begin{proof}}
\newcommand{\ethm}{\end{theorem}}
\newcommand{\elem}{\end{lemma}}
\newcommand{\epro}{\end{proposition}}
\newcommand{\edfn}{\bbox\end{definition}}
\newcommand{\erem}{\bbox\end{remark}}
\newcommand{\exam}{\bbox\end{example}}
\newcommand{\ecor}{\end{corollary}}
\newcommand{\eprf}{\end{proof}}
\newcommand{\beqn}{\begin{equation}}
\newcommand{\eeqn}{\end{equation}}
\newcommand{\bbox}{\vrule height7pt width4pt depth1pt}
\newenvironment{oldthm}[1]{\textsc{Theorem #1.} \em \noindent}{\par}
\newenvironment{oldlem}[1]{\par\noindent{\bf Lemma #1:} \em \noindent}{\par}
\newenvironment{oldcor}[1]{\par\noindent{\bf Corollary #1:} \em 
\noindent}{\par}
\newenvironment{oldpro}[1]{\par\noindent{\bf Proposition #1:} \em 
\noindent}{\par}
\newcommand{\othm}[1]{\begin{oldthm}{\ref{#1}}}
\newcommand{\eothm}{\end{oldthm} \medskip}
\newcommand{\olem}[1]{\begin{oldlem}{\ref{#1}}}
\newcommand{\eolem}{\end{oldlem} \medskip}
\newcommand{\ocor}[1]{\begin{oldcor}{\ref{#1}}}
\newcommand{\eocor}{\end{oldcor} \medskip}
\newcommand{\opro}[1]{\begin{oldpro}{\ref{#1}}}
\newcommand{\eopro}{\end{oldpro} \medskip}

\newcommand{\COMMENTOUT}[1]{}

\newcommand{\causal}[1]{\stackrel{#1}{\longrightarrow}}
\newcommand{\strandnext}{\Rightarrow}
\newcommand{\strandsend}{\rightarrow}

\newcommand{\<}{\langle}
\renewcommand{\>}{\rangle}
\newcommand{\cN}{\ensuremath{\mathcal{N}}}

\newcommand{\cC}{B}

\newcommand{\cA}{\ensuremath{\mathcal{A}}}

\newcommand{\cB}{\ensuremath{\mathcal{B}}}
\newcommand{\cR}{\ensuremath{\mathcal{R}}}
\newcommand{\R}{\ensuremath{\mathcal{R}}}

\newcommand{\cP}{\mathcal{P}}

\newcommand{\Hist}{\ensuremath{\mathit{hist}}}

\newcommand{\evt}{\ensuremath{\mathit{Evt}}}
\newcommand{\evts}{\ensuremath{\mathit{Evts}}}
\newcommand{\evtsseq}{\ensuremath{\mathit{EvtsSeq}}}
\newcommand{\seq}[2]{\ensuremath{\<#1\>_{#2}}}
\newcommand{\chains}{\ensuremath{\mathit{Chains}}}
\newcommand{\occ}{\mbox{occ}}

\newcommand{\joeheight}{height}

\newcommand{\MSG}{M}
\newcommand{\send}{\mbox{\textsf{send}}}
\newcommand{\sent}{\mbox{\textsf{sent}}}
\newcommand{\receive}{\mbox{\textsf{recv}}}
\newcommand{\nop}{\mbox{\textsf{no-op}}}
\newcommand{\term}{\mbox{term}}
\newcommand{\tr}{\mbox{tr}}
\newcommand{\Tr}{\mbox{Tr}}

\newcommand{\xor}{\mathit{Conf}}

\newcommand{\act}{\mathit{ACT}}
\newcommand{\sact}{\mathbf{a}}

\title{On the Relationship between Strand Spaces and Multi-Agent
Systems%
\thanks{
A preliminary version of this paper appears in the \emph{Proceedings
of the 8th ACM Conference on Computer and Communications Security},
2001.  Supported in part by NSF under grant IRI-96-25901 and 
IIS-0090145 and by ONR under grants  N00014-00-1-03-41 and
N00014-01-10-511, and by the DoD Multidisciplinary University Research
Initiative (MURI) program administered by the ONR under
grant N00014-01-1-0795. Authors' address: J. Y. Halpern, Department of
Computer Science, Cornell University, Ithaca, NY 14853,
email: \texttt{halpern@cs.cornell.edu}, home page:
\texttt{http://www.cs.cornell.edu/home/halpern},  R. Pucella, Department of
Computer Science, Cornell University, Ithaca, NY 14853,
email: \texttt{riccardo@cs.cornell.edu}.}
}
\author{JOSEPH Y. HALPERN and RICCARDO PUCELLA\\
Cornell University}
\begin{abstract}
\emph{Strand spaces} are a popular framework for the analysis of security
protocols. 
Strand spaces have some similarities to a formalism used successfully 
to model protocols for distributed systems, namely \emph{multi-agent
systems}. We explore the exact relationship between 
these two frameworks here.
It turns out that a key difference is the handling of
agents, which are unspecified in strand spaces and explicit in
multi-agent systems.  We provide a family of translations from strand
spaces to multi-agent systems parameterized by the choice of agents in
the strand space. We also show that not every multi-agent system of
interest can be expressed as a strand space. This reveals a lack of
expressiveness in the strand-space framework that can be
characterized by our translation. 
To highlight this lack of expressiveness, we show one simple way in
which strand spaces can be extended to model more systems.
\end{abstract}

\begin{document}

\maketitle

\section{Introduction}

\commentout{
\emph{Strand spaces} \cite{Thayer99} have recently emerged as a
popular framework for the analysis of security protocols. Roughly
speaking, the strand space corresponding to a protocol is the set of
the traces of the various interactions between the principals under 
consideration. 
Using strand spaces,
we can reason about the
secrecy of the values exchanged between principals and infer
authentication properties. The multi-agent
systems framework \cite{Fagin95} has 
been used to reason about distributed protocols,
and is based on a notion of runs; a run is a complete description of
what happens over time in one possible execution of the system.
While there are underlying similarities between the two approaches,
multi-agent systems seem more appropriate for reasoning about knowledge
and belief (an important component in reasoning about security;
cf.~\cite{BAN90}).  On the other hand, strand spaces have been used 
successfully to reason about security protocols.  As yet, multi-agent
systems have been less used
for that purpose
(although there have been initial attempts
to apply them to reasoning about cryptography and security;
cf.~\cite{HMT,Gray98}).  In any case, it seems worthwhile to
investigate 
the relationship between these two approaches, so as to attempt to
combine their strengths.
} %

\emph{Strand spaces} \cite{Thayer99} 
(THG from now on)
have recently emerged as a
popular framework for the analysis of security protocols. Roughly
speaking, the strand space corresponding to a protocol is the set of
the traces of the various interactions between the principals under 
consideration. Using strand spaces, we can reason about the
secrecy of the values exchanged between principals and infer
authentication properties. 
One limitation of the strand-space approach is that it assumes that
all the information available to a principal is either supplied
initially or contained in messages received by that principal.  
However, there is other important information that may also be available in a
security setting.  For example, an adversary may have information
about the protocol(s) being used.  Moreover, if the same agent is
playing different roles, then it may be able to combine information
it gathers in its various roles.
This information can be captured precisely using a formal model of
knowledge.  Indeed, the {\em multi-agent systems\/} framework 
used to represent the knowledge and belief of agents has been used quite
successfully to reason about distributed protocols (see \cite{Fagin95}
for intuition, details, and examples).  
This framework is
based on a notion of {\em runs}; a run is a complete description of what
happens over time in one possible execution of the system. 
Early attempts at applying the multi-agent systems framework to
reasoning about cryptography and security (cf.~\cite{Gray98,HMT})
suggest that these notions of knowledge and belief can be an important 
component in reasoning about security, the BAN logic being an example
in that particular direction \cite{BAN90}. Essentially, the idea is
simply that information can be derived in protocols not just through
the messages being exchanged, but through general properties of the
system. Our current project is to define a multi-agent systems framework
suitable for reasoning about security using notions such as knowledge
and belief. On the other hand, strand spaces have been used
successfully to reason about security protocols. Since there are
similarities between the two approaches, it is worthwhile to see how
much of the strand-space approach can be carried over to multi-agent
systems 
and vice versa.
This forces us to investigate in detail the relationship
between the two approaches. 
That is the purpose of this paper. 

\commentout{
We do that in this paper.  Given the close close relationship between
the two frameworks, perhaps not surprisingly, we 
provide a translation from strand spaces to  {\em strand systems}, a
subclass of multi-agent systems that seem to capture the intuition
underlying strand 
spaces.  In fact, we show that a strand space corresponds to more than
one strand system, depending on which notion of agents we impose on the
strand space.  
However, we show that there exist strand systems
which {\em cannot\/} be represented as strand spaces 
under any reasonable translation.  
Intuitively, there are causal
relationships which can be expressed in strand systems that simply
cannot be expressed in strand spaces.
This indicates a
fundamental lack of expressiveness in the current formulation of
strand spaces. 
Finally, we show how the strand-space framework can be extended so as to be
as expressive as strand systems.
} %

The key issue in relating the two frameworks is the
handling of agents. 
For our purposes, an agent is an entity (a principal, a process, etc.)
that can participate in interactions. 
This notion of agent is general enough to capture different
intuitions, depending on the kind of system being analyzed. Typically, 
an agent corresponds to a system-independent entity such as a
principal on behalf of whom interactions are performed. 
For our purposes, what matters is that an agent has
a state that is shared across all the interactions that the agent
performs.  
In 
multi-agent systems,
there is a clear notion of an agent 
participating in an interaction.
In 
strand spaces, there is not.
Each protocol interaction 
(described by a strand)
is viewed as independent
from all others.
In fact, each strand can be viewed as representing a different
agent. This approach to modeling agents  
is deliberate in the definition of strand spaces, and gives a theory
that yields general results.
Strand spaces do treat agents, in a fashion, by 
essentially assigning to every strand a name representing the ``agent''
executing the strand; see, for instance, the description of NSL spaces 
by THG
used to model the Needham-Shroeder-Lowe
protocol. However, it is still the case that strands corresponding to
the same ``agent'' can exchange values only through explicit
communication, i.e. there is no shared state across the strands
corresponding to the same ``agent'' name. For all intents and
purposes, these strands may as well be assigned to different actual
agents.

To highlight the role of agents, we 
provide a family of translations
from strand spaces to 
{\em strand systems}, a
subclass of multi-agent systems that seem to capture the intuition
underlying strand spaces.
The translations are 
parameterized by an assignment from strands
to agents.
This assignment associates with a strand the agent performing
the protocol interaction 
described by the strand.  Such an assignment captures 
the intuition that different strands can potentially be executed by the
same agent. 

Why is the role of the agents so significant? For the protocols
considered 
by THG,
it is not. On the other hand, it
is clear from the 
work on
BAN \cite{BAN90} and other logics 
(for instance, \cite{Stubblebine96,Syverson90}),
as well as the work on 
information flow \cite{McLean94a},
that 
belief and knowledge
are useful concepts when reasoning about security 
protocols. 
As we said earlier, there are a number of ways that an attacker can gain
knowledge in a system.  Certainly when an attacker intercepts a message,
it learns the contents of the message.  But it may learn much more if it
knows the protocol being run.  In addition, different principals
representing the same attacker may be able to pool the information they
have acquired.  In any case,
as soon as one talks about belief or knowledge, 
there must be agents in the picture to which 
belief or knowledge is ascribed. 
One advantage of a multi-agent system is that
it explicitly identifies agents and 
provides an easy way to 
ascribe
knowledge to agents (see \cite{Fagin95}).
In the context of security, that means we are forced to reason about, for
example, which principals represent the same agent or which ones
may represent the same agent.  (See \cite{Grove95,GroveH2} for logics
that carry out such reasoning explicitly and, in particular, distinguish
between agents and their names.)

Significantly, 
our translations are not surjective.  
Some strand systems
are not the image of any strand space, 
regardless of the 
assignment of agents to strands.
This is not just an artifact of our particular translation.  
Any translation from strand spaces to strand systems
that preserves 
the message history of the agents, in a precise sense, cannot be surjective.
Intuitively, this is because in 
a strand space
we cannot say ``either this sequence of events happens or that one
does, but not both''. 
This indicates a
fundamental lack of expressiveness in the current formulation of
strand spaces. 
One way to characterize this lack of expressiveness is by showing how
strand spaces can be extended to be able to model arbitrary strand
systems. We demonstrate one way of doing this by introducing a notion
of {\em conflict}, specifying when two strands 
cannot both be part of the same run.
We remark that 
the general
properties of strand spaces proved 
by THG,
such as the bounds on the penetrator,
are still valid in these extended strand spaces. 
We believe 
that
this notion of conflict becomes important when considering
modern security protocols.  Protocols such as SSL or TLS involve 
the selection of a subprotocol during the execution of a protocol
instance. Since only one such subprotocol can be chosen, it is natural 
to use conflict to model this. 

Despite this lack in expressiveness, strand spaces are quite
successful at analyzing typical protocols, particularly authentication 
protocols. Intuitively, based on the discussion above, this should be
due to those protocols not making any choice. We formalize this
intuition by exhibiting a property of protocols that ensures that a
strand system generated from a protocol with such a property (using
established techniques) is in fact the image of a strand space under
the natural translation. 

The rest of this paper is structured as follows. In Section
\ref{s:frameworks}, we review
strand spaces and
multi-agent systems. In Section \ref{s:tomas}, we present the 
translation from strand spaces to 
strand systems.
In Section \ref{s:frommas}, we discuss the problem of translating a 
strand
system into a strand space, and show why in
general we cannot perform the translation faithfully. In Section
\ref{s:expressiveness}, we describe an extension to the strand space
framework that is equivalent in expressive power to 
strand
systems. 
In Section~\ref{s:protocols}, we discuss the generation of systems
from protocols. 
We interpret our results in Section \ref{s:discussion}.
The proof of all technical results can be found in the Appendix.

\section{The frameworks}
\label{s:frameworks}

In this section, we review the two frameworks we want to relate, the
strand-space framework 
of THG,
and the multi-agent systems framework \cite{Fagin95}.

\subsection{Strand spaces}

Let $\MSG$ be the set of possible messages that can be exchanged by
the principals  in a protocol.\footnote{The actual contents of the message
and the 
structure of $\MSG$ are not important for the 
purpose of this paper.} A \emph{signed
term} is a pair $\<\sigma,u\>$ with 
$\sigma\in\{+,-\}$ and $u\in\MSG$. 
A signed term $\<+,u\>$ represents the sending of message $u$ and
is typically written $+u$, and a signed term $\<-,u\>$ represents
the reception of message $u$ and is typically written $-u$.
We write $(\pm\MSG)^*$ for the set of finite sequences of
signed terms. A \emph{strand space} over $\MSG$ 
consists of 
a set $\Sigma$,
whose elements are called \emph{strands}, 
together with a trace mapping $\mbox{tr}:\Sigma\rightarrow(\pm \MSG)^*$,
associating 
each strand in $\Sigma$ with
a sequence of signed terms. We typically
represent a strand space by the underlying set $\Sigma$, leaving the
trace mapping implicit.

In a strand space $\Sigma$, a \emph{node} is a pair $\<
s,i\>$, with $s\in\Sigma$ and an integer $i$ with $1\leq i\leq
|tr(s)|$. The set of nodes of $\Sigma$ is represented by $\cN$. We say 
the node $\< s,i\>$ \emph{belongs to} the strand $s$, and sometimes 
abuse notation by writing $\< s,i\>\in s$. Given a node
$n=\< s,i\>$, where $\mbox{tr}(s)=\<
\sigma_1,u_1\>\ldots\< \sigma_k,u_k\>$, define
$\mbox{term}(n) = \< \sigma_i,u_i\>$. If $n_1$ and
$n_2$ are nodes, we write $n_1\strandsend n_2$ if $\mbox{term}(n_1)=+u$
and $\mbox{term}(n_2)=-u$; 
we write $n_1\strandnext n_2$ if both $n_1$
and $n_2$ occur on the same strand $s$ and 
$n_1=\< s,i\>$ and 
$n_2=\< s,i+1\>$. Note that the set $\cN$ of nodes 
together
with both sets of
edges $n_1\strandsend n_2$ and $n_1\strandnext n_2$ 
forms a directed
graph $(\cN,(\strandsend\cup\strandnext))$.

Strand spaces are aimed at reasoning about 
the
security of systems in the
presence of a hostile penetrator with various capabilities. In order to model
such a 
penetrator, a notion of an
{\em infiltrated strand space\/}
is defined
by THG;
the infiltrated strand space
contains both regular strands and a
set of so-called \emph{penetrator strands} that represent the actions
available to a penetrator. For the purposes of this paper, there is no 
need to distinguish penetrator strands from regular
strands, so we do not consider infiltrated strand spaces.

A bundle 
represents
a snapshot of a possible protocol  
execution. For a given strand space $\Sigma$, let
$\cC = (\cN_{\cC},(\strandsend_{\cC}\cup\strandnext_{\cC}))$ be a
subgraph of $(\cN,(\strandsend\cup\strandnext))$. 
The graph
$\cC$ is a \emph{bundle} if
\begin{enumerate}
\item[B1.] $\cC$ is finite,
\item[B2.] if $n_2\in\cN_{\cC}$ and $\mbox{term}(n_2)$ is negative, then
there is a unique $n_1$ such that $n_1\strandsend_{\cC} n_2$,
\item[B3.] if $n_2\in\cN_{\cC}$ and $n_1\strandnext n_2$, then
$n_1\strandnext_{\cC} n_2$,
\item[B4.] $\cC$ is acyclic.
\end{enumerate}
In B2 and B3, 
because $\cC$ is a graph, it follows that 
$n_1\in\cN_{\cC}$. We say a node $n$ is in the bundle $\cC$ if it is
in $\cN_{\cC}$. 

It will be useful for us in this paper to allow infinite bundles.
An {\em infinite bundle\/} is just a subgraph of 
$(\cN,(\strandsend\cup\strandnext))$ that satisfies B2--4 
(that is, we no longer require the 
finiteness
condition B1).
The {\em \joeheight{}\/} of an infinite bundle is the length of 
the 
longest finite sequence of nodes $n_1, n_2, n_3, \ldots, n_k$ in $\cC$ such 
that
$n_1\leadsto n_2 \leadsto \ldots \leadsto n_k$,
where $\leadsto$ is either
$\strandsend$ or $\strandnext$.  (A bundle can have infinite \joeheight{} if
there is no bound on the length of the longest sequence of this type.)
Of course, all finite bundles have finite \joeheight{}.  
It is easy, however,
to construct infinite bundles of infinite \joeheight{} (even if all individual
strands have length at most 2).  For example, 
consider the strand space $\Sigma=\{s_i : i\in\mathbb{Z}\}$, with a
trace mapping $\tr(s_i)=\< -u_i,+u_{i+1}\>$. The 
strand space $\Sigma$ itself in this case is an infinite bundle of infinite 
\joeheight{}.
All the arguments of 
THG
which were applied to finite 
bundles 
go through without change for 
infinite
bundles
of finite \joeheight{}.  (Indeed, they go through for infinite bundles that
are {\em well-founded}, in the sense of having no infinite
``descending'' sequences of the form  
$\ldots \leadsto n_3
\leadsto n_2 \leadsto n_1$, although we end up using only bundles of
finite \joeheight{} in our arguments.)

\newcommand{\cheight}[1]{\ensuremath{#1\mbox{-height}}}
\newcommand{\ctrace}[1]{\ensuremath{#1\mbox{-trace}}}
\commentout{
If $\cC$ is a bundle, we define the {\em \cheight{\cC}\/} of a strand $s$ to
be the largest $i$ such that $\< s,i\>\in\cN_{\cC}$. We
define $\ctrace{\cC}=\<
\mbox{tr}(s)_1,\ldots,\mbox{tr}(s)_m\>$ where
$m=\cheight{\cC}(s)$. 
We often write $\crestr{s}{\cC}$ for $\ctrace{\cC}(s)$
and 
$\csize{\crestr{s}{\cC}}$ for $\cheight{\cC} (s)$ (intuitively, the
length of $s$ in $\cC$). We write $s\in\cC$ if all
the nodes in a strand $s$ are in the bundle $\cC$, and $s\preceq\cC$
if at least one node of $s$ is in $\cC$ (equivalently, if
$\csize{\crestr{s}{\cC}}>0$). 
}

\subsection{Multi-agent systems}

In the multi-agent systems approach, every agent is assumed to be in
some \emph{local state} at each point in time.
Given a set $\cA$ of agents,
we characterize a system over $\cA$ at a given
point in time in terms of a \emph{global state}; this is a 
tuple $\<\sigma_a : a\in\cA\>$, where $\sigma_a$
is the local state of agent $a$. The
local states of an agent intuitively encode all the information that
the agent has available at a given point in time. 
In typical distributed systems applications, 
the local state includes
the values of variables and a history of messages
received. 
If we are modeling a group of agents
playing a poker game, the local state 
may
include the cards that the
agent holds and the bets that have been made thus far. 

To capture changes to the system over time, 
we define a 
{\em run\/} of the system to be a
function from time to global states. Intuitively, a run is a complete
description of what happens over time in one possible execution of the 
system. A 
{\em point\/} is a pair $(r,m)$ consisting of a run $r$ and a time
$m$. The global state $r(m)$ describes the state of the system at the
point $(r,m)$. Formally, we take a \emph{system} to consist of a set of
runs. 
Informally,
the system includes all the possible executions of
the system, that is, all the different ways it could evolve through
time. 

Due to the assumptions made by the strand-space approach, 
namely that events in strands consist of sending and receiving
messages, we 
consider only systems where the 
local state of an agent is the sequences of messages 
that
the agent has sent and received. 
Thus, we  
deliberately
ignore internal actions (or, more accurately,
treat them as irrelevant).

We can formalize the above description as follows. Consider a fixed
set $\MSG$ of messages.
A {\em history\/} for 
agent
~$a$ (over 
$\MSG$) is a sequence 
of elements of the form $\sent(u)$ and
$\receive(u)$, where~$u \in \MSG$.
We think of $\sent(u)$ as representing the event ``message
$u$ is sent'' and $\receive(u)$
as representing the event ``message~$u$ is received.''
Intuitively, $a$'s history at $(r,m)$ consists of~$a$'s initial state, 
which we take to be 
the empty sequence,
followed by the sequence describing~$a$'s actions up to time~$m$.
If~$a$ performs no
actions in round~$m$, then its history at $(r,m)$ is the same as its
history at $(r,m-1)$.%
\footnote{Round $m$ takes place between time $m-1$ and time $m$.  
Actions are performed during a round.  The effect of an action performed
by agent $a$ at round $m$ appears in agent $a$'s state at time $m$.}
In such a \emph{message-passing system}, we speak
of $\sent(u)$ and $\receive(u)$
as {\em events}.
For $a \in \cA$, let $r_a(m)$ be agent $a$'s history in $(r,m)$.
We say that an event $e$ {\em occurs
in $a$'s history
in round $m+1$ of run~$r$\/}
if $e$ is in (the sequence) $r_a(m+1)$ but not in $r_a(m)$.

In a message-passing system, the 
agent's
local state at any point
is its history.  Of course, if~$h$ is the history of agent~$a$ at the
point $(r,m)$, then we want it to be the case that~$h$ describes what
happened in~$r$ up to time~$m$ from~$a$'s point of view.   To do this,
we need to impose some consistency conditions on global states.
In particular, we want to ensure that message
histories 
do not shrink
over time, and that every
message
received in round~$m$ corresponds to a message that was sent 
at some earlier round.
Given
a set $\MSG$ of messages,
we define a {\em message-passing system\/}
(over $\MSG$) to be a system such that for each point $(r,m)$
and each agent $a \in \cA$,
the following constraints are satisfied:
\begin{itemize}
\item[MP1.] $r_a(m)$ is a history over 
$\MSG$;
\item[MP2.]
for every event $\receive(u)$ in $r_a(m)$ there exists a
corresponding event $\sent(u)$ in $r_b(m)$,  for some $b \in
\cA$;\footnote{To simplify our translations, we allow an agent to send
a message to itself, so $a$ and $b$ can be the same agent.} 
\item[MP3.]
$r_a(0)$ is the empty sequence
and $r_a(m+1)$ is either identical to $r_a(m)$ or the result of
appending one event to $r_a(m)$.
\end{itemize}
MP1 says that 
an agent's
local state
is its history, MP2 guarantees that every message
received at round~$m$ corresponds to one that was sent 
earlier, and
MP3 guarantees that histories do not shrink.

We think of strand spaces as completely asynchronous message-passing
systems. Roughly speaking, strand spaces do not place any
constraints on the relative order
of events in different agents' histories beyond those imposed by MP1 and
MP2.  As argued in \cite[Section 4.4.6]{Fagin95}, we can capture such
asynchrony by considering systems that consist of {\em all\/} runs
satisfying MP1--3 for some set of histories.  
Formally, we say that 
$\cR$ is a \emph{strand system} if there exists a 
sequence $\<V_a: a \in  \cA\>$, where $V_a$ is a set of 
histories
over 
some set $\MSG$ of messages,
such that $\cR$ consists of all runs
satisfying MP1--3 
where
agent $a$'s local state is a history in
$V_a$ at every point.
We call $\cR$ the strand system {\em generated by\/} $\<V_a: a \in  \cA\>$.
Informally,
the set $V_a$ specifies  the possible histories agent $a$
could have.  The strand system generated by $\<V_a: a \in  \cA\>$
consists of
all runs satisfying MP1--3 such that agent $a$'s histories are in $V_a$
for all $a \in \cA$.

Strand systems are closely related to the
\emph{asynchronous message-passing systems (amps)} defined in
\cite[Chapter 4]{Fagin95}.  
The main difference is that for strand systems, messages
are anonymous. A message does not specify a sender or a
receiver.  Messages in amps, on the other hand, are not anonymous.
Events have the form $\sent(u,a,b)$ ($u$ is sent to $a$ by $b$) and
$\receive(u,a,b)$ ($u$ is received by $b$ from $a$).  
The remaining differences are minor.
Strand systems allow
for an infinite number of agents, whereas in amps there are only
finitely many agents. Amps can be easily modified 
so as to allow infinitely many agents. 
Moreover, agents are allowed in amps to have a nontrivial initial state, while
for strand systems, the initial state is always the empty 
sequence.  This was done for compatibility with the definitions 
of THG.
\section{Translating strand spaces to strand systems}
\label{s:tomas}

In this section, we 
consider
the problem of translating strand spaces
into strand systems. We do this by formalizing the strand space
intuition that bundles represent snapshots of possible executions. Our 
construction derives the possible execution traces in terms of
sequences of bundles, which are then used to construct the runs of the
system. 

A multi-agent system requires an explicit set of agents; a strand space
does not. To perform the translation, we specify a set $\cA$ of
agents and a particular \emph{agent assignment} 
$A: \Sigma \rightarrow \cA$, which intuitively associates with each
strand $s \in \Sigma$ the agent $A(s)$ executing $s$.
In the generated strand system, an agent behaves
as if it were concurrently executing the various strands
assigned to it. The 
motivation
behind this approach is that if the same 
agent is in reality executing many strands, then it should
share its knowledge across all the strands it is executing. 

The choice of agents and the agent assignment for a given strand space 
is 
left to the model designer.
Different choices lead to different multi-agent systems. As
we show at the end of this section, 
associating a different agent
with each strand
enforces 
the basic strand space tenet that 
information is exchanged
only through explicit messages,
i.e.~there is no shared state between different strands.

The translation takes as arguments
a strand space $\Sigma$, a set $\cA$ of agents, and an agent
assignment $A$ from strands in $\Sigma$ to agents.  
To define the translation, we first define a
relation on bundles 
that
represents the actions
that the agents in the strand space can 
perform.
Given a strand $s \in \Sigma$ and a bundle $\cC$, let 
$\cheight{B}(s)$
be the largest $i$ such that $\<
s,i\>\in\cN_{\cC}$.  (We take 
$\cheight{B}(s)
= 0$ if no
node in $s$ appears in 
$\cC$.)\footnote{This notion of height of a strand in a bundle should
not be confused with the notion of height of a bundle we defined in the
previous section.}
A function $f:\Sigma\rightarrow\Sigma$ \emph{respects $A$} if
$A(s)=A(f(s))$, that is, the same agent is associated with both
strands $s$ and $f(s)$ for all strands $s \in \Sigma$.
If $B_1,B_2$ are
(possibly infinite)
bundles of $\Sigma$,
and $f: \Sigma \rightarrow \Sigma$ is a bijection that 
respects $A$,
we 
write
$B_1 \sqsubseteq_f B_2$ if the following two conditions hold:
\begin{enumerate}
\item if $\< s,i \>$ is in $B_1$, then $\< f(s),i\>$
is in $B_2$ and 
$\term(\<s,i\>) = \term(\<f(s),i\>)$,
\item if $\< s,i \> \strandsend \< s',j \>$ is an
edge in $B_1$, then $\< f(s),i \> \strandsend \< f(s'),j
\>$ is an edge in $B_2$.
\end{enumerate}
These clauses guarantee that the prefix of $s$ that is in $B_1$ is
a prefix of the prefix of $f(s)$ that is in $B_2$.  
For example, 
if $B_1$ consists of the single node $\<s,1\>$ and $B_2$ consists of
$\<s',1\>$ and $\<s',2\>$, where $\term(\<s,1\>) = \term(\<s',1\>)$, then
$B_1 \sqsubseteq_f B_2$, where $f$ is the bijection that permutes $s$ and
$s'$, while acting as the identity on all other strands.

For many cases of interest, we can simply take the bijection $f$ to be
the identity; in that case, $B_1 \sqsubseteq_f B_2$ 
if and only if
$B_1$ is a subgraph of $B_2$.  We discuss the reason for allowing arbitrary
bijections 
and the role of the bijection at the end of this section.

We write $B_1\mapsto B_2$ if there is a bijection $f: \Sigma \rightarrow
\Sigma$ 
that respects $A$ such that
\begin{enumerate}
\item $B_1\sqsubseteq_f B_2$, and
\item 
$\sum_{s\in A^{-1}(a)}
\cheight{B_2}(f(s))-\cheight{B_1}(s)
\leq1$
for all agents $a\in\cA$.
\end{enumerate}
Informally, $B_1\mapsto B_2$ 
if, 
for each agent $a \in \cA$, 
$B_2$ extends the prefix of at
most one strand in $B_1$ corresponding to $a$, and extends it by at most
one node.  
(Note that the strand $f(s)$ in $B_2$ extending the prefix of strand $s$
in $B_1$ may be different from $s$, depending on the definition of $f$.)
If $B_2$ does extend the prefix of one of the strands in $B_1$ corresponding to
agent $a$ by one node, let $e_{a,B_1 \mapsto B_2}$ denote the event
corresponding to that 
node: 
if the node is $n$ and $\term(n) =
+u$, then $e_{a,B_1 \mapsto B_2}$ is $\sent(u)$,
and if $\term(n) =
-u$, then $e_{a,B_1 \mapsto B_2}$ is $\receive(u)$.
We define a 
$\mapsto$-\emph{chain} (or simply a chain) to be 
an infinite sequence 
of bundles $B_0,B_1,\ldots$ such that
$B_0$ is the empty bundle and $B_0\mapsto B_1\mapsto\ldots$. 
Let $\chains(\Sigma,\cA,A)$ be the set of all 
chains in $\Sigma$. 
We associate with every chain
in $\chains(\Sigma,\cA,A)$ a 
run as follows:
Given a chain $C = B_0 \mapsto B_1 \mapsto \ldots$ and an agent $a \in
\cA$, define $\Hist_a^m(C)$ inductively. 
Let $\Hist_a^0(C) = \< \, \>$; let  $\Hist_a^{n+1}(C) =
\Hist_a^n(C)$ 
if no strand corresponding to agent $a$ in $B_n$ is
extended in $B_{n+1}$; otherwise, let $\Hist_a^{n+1}(C) = 
\Hist_a^n(C) \cdot  e_{a,B_n \mapsto B_{n+1}}$. 
(Informally,
$\Hist_a^{n+1}(C)$ is the result of appending to $\Hist_a^n(C)$ the
unique event performed by agent $a$ in going from $B_n$ to $B_{n+1}$.)
Thus, $\Hist_a^{n}(C)$ consists of all the events that $a$ has performed
in $B_n$.
Let $r^C$ be the run such that $r^C_a(m) = \Hist_a^m(C)$ and let
$\R(\Sigma,\cA,A) = \{r^C: C \in \chains(\Sigma,\cA,A)\}$.

\thm\label{t:general} $\R(\Sigma,\cA,A)$ is a strand 
system. \ethm

In light of Theorem~\ref{t:general}, define the map $T_A$ from strand
spaces to strand systems by taking $T_A(\Sigma) = \R(\Sigma,\cA,A)$.

As we mentioned at the beginning of this section, we can model strand 
spaces as 
discussed
by THG
by taking the set of agents of a strand space $\Sigma$ to be $\Sigma$,
and taking the identity function $\mathit{id}$ as the agent
assignment. This captures explicitly the intuition that strands are
independent protocol executions, 
that for all intents and purposes
may be assumed to be executed by different agents. 
This is the case since there is no state shared between strands, and
every communication is made explicit. In other words, there is no
conceptual difference between two strands $s_1$ and $s_2$ executed by
different processes of an agent or by two distinct agents if there
cannot be any shared state between $s_1$ and $s_2$.

There is a small amount of information that is lost in the
translation from strand spaces to strand systems, which will become
evident in Theorem~\ref{t:adequate} below. This loss stems from the fact
that messages in strand systems are completely anonymous. 
For example, 
if agent 2 and agent 3 both send a message $u$ and later agent 1
receives it, there is no way in a strand system to tell if agent 1
received $u$ from agent 2 or agent 3.  By way of contrast, in a strand
space, there is an edge indicating who agent 1 received the message
from.  
The multi-agent system framework can in fact 
keep track of who an agent received a message from 
by adding an additional component to the global
state; this is the state of the \emph{environment}, which intuitively
describes everything relevant to the system not included in the local
states of the processes.\footnote{In our particular case, the
environment could record the sender of each message that is 
received at any given round.} We will not bother going into the
details of the environment in this paper, as the issue 
does not affect our results. We can 
characterize the 
information loss resulting from our translation
by defining a relation between globals states of
$\cR(\Sigma,\Sigma,\mathit{id})$ and bundles of $\Sigma$. We say that
a global state $\<\sigma_s :~ s\in\Sigma\>$  (recall that 
here
$\cA=\Sigma$)
is \emph{message-equivalent} to a bundle $B$ if for each $s\in\Sigma$,
if $\sigma_s = \<e_1,\ldots,e_k\>$ then $\cheight{B}(s)=k$ and, for each
$i$ such that $1\leq i\leq k$, if $\term(\<s,i\>)=+u$ then $e_i$ is
$\sent(u)$, and if $\term(\<s,i\>)=-u$ then $e_i$ is $\receive(u)$. 
Intuitively, a global state is message-equivalent to any bundle that has the
same nodes. This captures the intuition that an agent receiving a
message is not aware of the sender. The following theorem shows that,
except for this loss of information, our translation from strand
spaces to strand systems essentially identifies bundles and global
states (if we treat all strands as being associated with a different agent).

\thm\label{t:adequate}
Every global state of $\cR(\Sigma,\Sigma,\mathit{id})$ is message-equivalent
to a bundle of $\Sigma$ of finite \joeheight{}, and every bundle of
$\Sigma$ of finite \joeheight{} is message-equivalent to a global state of
$\cR(\Sigma,\Sigma,\mathit{id})$.
\ethm
We remark 
that if the environment state is used
to record the sender of each 
received message, Theorem~\ref{t:adequate} can be strengthened to a
1-1 correspondence between global states of
$\cR(\Sigma,\Sigma,\mathit{id})$ and bundles of $\Sigma$ of finite 
\joeheight{}. 

With these results in hand, we now discuss some of the choices 
made, in particular, why we allowed infinitely many agents, infinite
bundles, and an arbitrary bijection $f$ in the definition of
$\mapsto$.  It turns out that these choices are
somewhat related.  First observe that, in 
Theorem~\ref{t:adequate},
we identified each strand with an agent.  Thus, if there are infinitely
many strands in the strand space, 
the corresponding strand system requires
infinitely many agents.  
Naturally, if we restrict our analysis
to strand spaces with only finitely many strands,
then 
we can 
take the corresponding 
strand systems 
to have only 
finitely many agents.  Infinite bundles are needed in order to prove 
Theorem~\ref{t:general} 
when
there are infinitely many agents.  To
understand why, consider a strand space $\Sigma$, where $\Sigma =
\{s_1, s_2, \ldots \}$ and $\tr(s_n) = \< +u_n\>$.  
In other words, strand $s_n$
has exactly one node, at which a send action is performed.
If a different agent is associated with each strand, then in the
corresponding strand system,
the set of histories for agent $n$ will
consist of the empty history and the history $\<\sent(u_n)\>$.
The system based on this set of histories has a run where all the agents
send their message simultaneously at round 1.  This history corresponds
to the infinite bundle consisting of all the strands in $\Sigma$.  
Intuitively, if all the agents can send a message, there is no reason
that they should not all send it 
the first round.
Why do strand spaces allow infinitely many 
strands? 
Often, security
protocols rely on \emph{nonce values}, which are values guaranteed to
be unique within a run 
of the system. Strand spaces model nonce values 
by specifying a different strand for each possible value of a
nonce. Since, theoretically, there can be infinitely many nonces (as a
consequence of uniqueness), we typically have to consider infinitely
many strands for a given protocol. 
Note that these strands do not
necessarily  represent computations of {\em different\/} agents.
Indeed, it probably makes sense to consider them all as being performed
by the same agent (but at most one of them being performed in a given
execution of the protocol).  

The bijection $f$ in $\sqsubseteq_f$
is not needed if a different agent is associated with
each strand.  (That is, in this case it suffices to take $f$ to be the
identity.)  Similarly, $f$ is not needed if there is a bound $k$ on the
length of all strands in $\Sigma$.  Indeed, it is needed only to take
care of the possibility that there is an infinite sequence of strands,
each intuitively a prefix of the next, and all associated with the same
agent.  For example, consider the strand space  $\Sigma$ where,
again, $\Sigma = \{s_1, s_2, \ldots \}$ but now $\tr(s_n) = \<+u_1,
\ldots, +u_n\>$.  Intuitively, in this strand space, $s_n$ is a
substrand of $s_{n+1}$ (although, formally, there is no notion of
substrand in strand spaces).  Suppose 
that
the mapping is such that $\cA$
consists of one agent $a_1$ and $A$ associates all the strands in
$\Sigma$ with $a_1$.  
If we did not allow such a map $f$ (or, equivalently, required $f$ to be
the identity), then the only chains would be those of the form
$B_0\mapsto B_1\mapsto
\ldots\mapsto B_k\mapsto B_k\mapsto B_k\mapsto\ldots$ (for some finite  
$k$), where, for some strand $s$, each $B_i$ is a prefix of $s$.
If we apply our mapping to this collection of strands, in the resulting
system, there is a single set of histories $V_{a_1} =
\{\<\sent(u_1)\>,\<\sent(u_1),\sent(u_2)\>,\<\sent(u_1),\sent(u_2),\sent(u_3)\>,\ldots\}$, 
where
each history in $V_{a_1}$ is finite. However, 
the system generated by this set of histories contains an infinite
run, which sends message $u_i$ at time $i$. 
Unfortunately, there is no chain
corresponding to this run. 
On the other hand, once we allow nontrivial bijections $f$, 
there is no problem.
Abusing notation somewhat, there is a chain of the form $s_1 \mapsto
s_2 \mapsto s_3 \mapsto \ldots $ where $a_1$'s history is unbounded,
since $s_k \sqsubseteq_{f_k} s_{k+1}$, where $f_k$ permutes $s_k$ 
and $s_{k+1}$ and is the identity on all other strands.

Intuitively, 
if $f$ must be the identity, then
every chain must ``choose'' the
strand it is executing, which implicitly corresponds to choosing how
many messages to send in that particular run. 
By providing a function
$f$ that permits us to ``jump'' to strands with the same prefix between 
any consecutive bundles of a chain,
we are essentially modeling an agent that does not choose the length of
the strand up front, but rather just performs the actions (and thus, if
one strand is a prefix of another, it cannot tell which of the two
strands it is performing).

While it is important to recognize these subtleties, they do not arise 
in most protocols. 
For instance, strands for specific protocols will typically
be of bounded length, and therefore the bijection $f$ is not  needed
to define chains in the corresponding strand space.

\section{Translating strand systems to strand spaces}
\label{s:frommas}

In this section, we consider 
the translation of strand systems into strand spaces.  
Specifically,
given a strand system $\cR$, 
is there 
a strand space which maps 
to $\cR$ under a suitable agent assignment? 
In general, there is not. 
This result is not an artifact of our 
translation,
but reflects a fundamental difference between strand spaces and strand 
systems.
In particular, it does not depend on 
any of 
the subtleties that were pointed
out at the end of last section. 

To understand the difficulties,
consider the following simple system $\cR_1$.
It essentially contains two runs
$r_1$ and $r_2$, with distinct messages $x,y,u,v$:

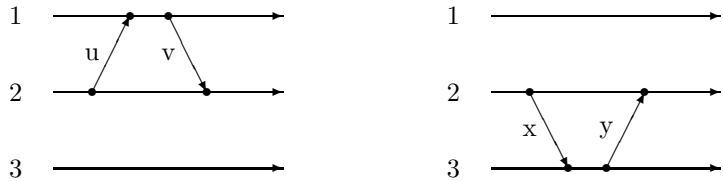
\begin{center}
\begin{picture}(6,4) %
\put(1,0){\vector(1,0){6}}
\put(1,2){\vector(1,0){6}}
\put(1,4){\vector(1,0){6}}
\put(0,0){\makebox(0,0){3}}
\put(0,2){\makebox(0,0){2}}
\put(0,4){\makebox(0,0){1}}
\put(2,2){\circle*{.2}}
\put(3,4){\circle*{.2}}
\put(2,2){\vector(1,2){1}}
\put(2,3){\makebox(0,0){u}}
\put(4,4){\circle*{.2}}
\put(5,2){\circle*{.2}}
\put(4,4){\vector(1,-2){1}}
\put(4,3){\makebox(0,0){v}}
\end{picture} 
\hspace{1in} 
\begin{picture}(6,4) %
\put(1,0){\vector(1,0){6}}
\put(1,2){\vector(1,0){6}}
\put(1,4){\vector(1,0){6}}
\put(0,0){\makebox(0,0){3}}
\put(0,2){\makebox(0,0){2}}
\put(0,4){\makebox(0,0){1}}
\put(2,2){\circle*{.2}}
\put(3,0){\circle*{.2}}
\put(2,2){\vector(1,-2){1}}
\put(2,1){\makebox(0,0){x}}
\put(4,0){\circle*{.2}}
\put(5,2){\circle*{.2}}
\put(4,0){\vector(1,2){1}}
\put(4,1){\makebox(0,0){y}}
\end{picture}
\end{center}
Because 
the MP1--3 assumptions on strand systems
allow arbitrary delays between the events,
there are more than two runs in the system;
the essential fact is that, in any given run, agent $2$
communicates only with agent $1$ or only with agent $3$.
Formally, $\cR_1$ is the strand system generated by taking:
\[\begin{array}{l}
V_1 = \{\<\>,\<\receive(u)\>,\<\receive(u),\sent(v)\>\},\\
V_2 = \{\<\>,\<\sent(u)\>,\<\sent(x)\>,\<\sent(u),\receive(v)\>
,\<\sent(x),\receive(y)\>\}, \mbox{and}\\
V_3 = \{ \<\>,\<\receive(x)\>,\<\receive(x),\sent(y)\>\}.\end{array}\]

Under the mapping presented in the previous section, there does not exist
a strand space that maps to this system, for any agent assignment.
Intuitively, any strand space modeling the
system $\cR_1$ will need at least strands corresponding to runs $r_1$
and strands corresponding to runs $r_2$. Since these sets of strands
do not interact (that is, they do not exchange any message),  the
translation of Section \ref{s:tomas} will produce  a system 
that contains runs that amount to all possible 
interleaving of the strands corresponding to $r_1$ and $r_2$.
This results in a system that is strictly larger than $\cR_1$.  For
example, it must contain runs with the following histories for
agents 1, 2, and 3:

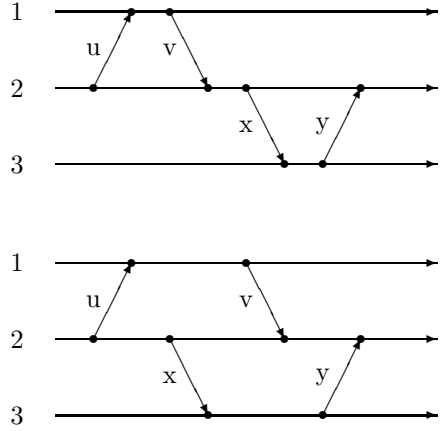
\begin{center}
\begin{picture}(10,4) %
\put(1,0){\vector(1,0){10}}
\put(1,2){\vector(1,0){10}}
\put(1,4){\vector(1,0){10}}
\put(0,0){\makebox(0,0){3}}
\put(0,2){\makebox(0,0){2}}
\put(0,4){\makebox(0,0){1}}
\put(2,2){\circle*{.2}}
\put(3,4){\circle*{.2}}
\put(2,2){\vector(1,2){1}}
\put(2,3){\makebox(0,0){u}}
\put(4,4){\circle*{.2}}
\put(5,2){\circle*{.2}}
\put(4,4){\vector(1,-2){1}}
\put(4,3){\makebox(0,0){v}}
\put(6,2){\circle*{.2}}
\put(7,0){\circle*{.2}}
\put(6,2){\vector(1,-2){1}}
\put(6,1){\makebox(0,0){x}}
\put(8,0){\circle*{.2}}
\put(9,2){\circle*{.2}}
\put(8,0){\vector(1,2){1}}
\put(8,1){\makebox(0,0){y}}
\end{picture}

\vspace{.5in}
\begin{picture}(10,4) %
\put(1,0){\vector(1,0){10}}
\put(1,2){\vector(1,0){10}}
\put(1,4){\vector(1,0){10}}
\put(0,0){\makebox(0,0){3}}
\put(0,2){\makebox(0,0){2}}
\put(0,4){\makebox(0,0){1}}
\put(2,2){\circle*{.2}}
\put(3,4){\circle*{.2}}
\put(2,2){\vector(1,2){1}}
\put(2,3){\makebox(0,0){u}}
\put(6,4){\circle*{.2}}
\put(7,2){\circle*{.2}}
\put(6,4){\vector(1,-2){1}}
\put(6,3){\makebox(0,0){v}}
\put(4,2){\circle*{.2}}
\put(5,0){\circle*{.2}}
\put(4,2){\vector(1,-2){1}}
\put(4,1){\makebox(0,0){x}}
\put(8,0){\circle*{.2}}
\put(9,2){\circle*{.2}}
\put(8,0){\vector(1,2){1}}
\put(8,1){\makebox(0,0){y}}
\end{picture}
\end{center}

Roughly speaking, what is happening in the strand system is that agent~2
nondeterministically decides whether to send message $u$ to agent~1 or
message~$x$ to agent~3.  In any run of the system, it sends one or the
other, but not both.  The problem here is that, in the strand-space
framework, we cannot say ``one or the other, but not both''.

To make this precise, given an agent assignment $A$, define a translation
$T$ from strand spaces to strand systems to be {\em $A$-history
preserving\/} if, given a strand space $\Sigma$,
\begin{itemize}
\item for each agent $a \in \cA$, run $r \in T(\Sigma)$, and time
$m$, there exists a bundle $\cC$ 
in $\Sigma$
such that the  events in agent $a$'s history $r_a(m)$ are exactly
those that appear in nodes $\< s, i \>$ in $\cC$ such that $A(s) = a$; 
\item conversely, for each agent $a \in \cA$ and bundle $\cC$
of finite \joeheight{}
in $\Sigma$, there exists a run $r \in T(\Sigma)$ and time
$m$ such that the
events in agent $a$'s history $r_a(m)$ are exactly those that appear in
nodes $\< s, i \>$ in $\cC$ such that $A(s) = a$.
\end{itemize}

Notice that the translation $T_A$ defined in the previous section is
$A$-history preserving. 

\thm\label{t:notinimage}
There is no agent assignment $A$ and 
$A$-history preserving translation $T$ from strand spaces to
strand systems such that the strand system $\cR_1$ is in the image of $T$.
\ethm
The example above suggests that in general, systems arising from an
agent running a nondeterministic protocol may not be the image of a
strand space under our translation. The problem in fact is more
profound. Even if the agents are running deterministic protocols, the
nondeterminism inherent in the delay of messages delivery may prevent
a system from being the image of a strand space. Consider the
following system with two agents. Agent 1 sends a message $u$ to agent 
2. If agent 2 hasn't received it yet, and hasn't sent a
$\mathit{nack}$ message yet, she sends a $\mathit{nack}$. When she
gets message $u$, she sends an $\mathit{ack}$. Here, the strand space
intuitively corresponding to this situation will include a strand for
agent 1 where he sends $u$. For agent 2, we can consider at least the
following two strands, $\<-u,+\mathit{ack}\>$ and
$\<+\mathit{nack},-u,+\mathit{ack}\>$. One can check that there exists 
a chain leading to the bundle made up of the following strand
prefixes: $\<+u\>$, $\<-u,+\mathit{ack}\>$, and $\<+\mathit{nack}\>$,
leading, through our translation, to a possible history for agent
agent of the form
$\<\receive(u),\sent(\mathit{ack}),\sent(\mathit{nack})\>$, which does 
not arise in the original system. In this example, the problem does
not occur because the agent makes a choice, but, intuitively, because
the ``environment'' is making a choice when delivering messages. We
will revisit these issues in Section~\ref{s:protocols}, when we study
the generation of systems from protocols.

\section{Extended strand spaces}
\label{s:expressiveness}

In the previous section, we showed that not all strand systems 
correspond to strand spaces.  More precisely, we showed that some
strand spaces could not be in the image of any history-preserving translation.
How reasonable is the requirement that a translation be history
preserving?  Suppose that $T$ is a translation from strand spaces to
strand systems that is ``acceptable'' in some sense.  It certainly seems
reasonable to require that if $T(\Sigma) = \R$, then the events in
every history $r_a(m)$ that arises in $\R$ correspond to events that
agent $a$ actually performed in some bundle.  (Note that there is no
need to consider infinite bundles here; if there is a bundle at all, it
is finite.)  Conversely, given a bundle $\cC$ over $\Sigma$, it seems
reasonable to require that there exists a history where $a$ performs the
same actions as it does in the bundle.  

So exactly why is there no strand space corresponding to the system
$\cR_1$?  Roughly speaking, given a strand space 
$\Sigma$, any set of strands that satisfies B1--4 is a bundle.
Thus, once certain bundles exist, others are forced to exist too,
including ones that do not correspond to any run in $\cR_1$.  
For example,
once 
there is 
a bundle corresponding to ``2 sends $u$ to 1 and
gets a response $v$'', and another bundle corresponding to ``2 sends $x$
to 
3
and gets a response $y$'', 
there has to be 
a bundle where 2 both sends $u$ to 1 and sends 
$x$
to 3.  
The strand-space framework cannot express 
``either this sequence of events happens or that one does, but not 
both''.  
As we now show, this is essentially the only impediment standing in the
way of a translation from strand spaces to strand systems.
We extend the strand-space formalism with a notion of conflict 
that allows us to 
prohibit
certain strands from appearing together in the same bundle,
and then show that such extended strand spaces can model all strand
systems.\footnote{We do not want to imply that this is the only way to 
extend strand spaces to achieve this effect, nor do we claim that this 
approach is particularly original. Indeed, there is a vast
literature in concurrency theory on the subject of implementing choice 
constructs in various formalisms; see, for instance, 
\cite{Busi94,Palamidessi97}.
Independently, Crazzolara and Winskel \citeyear{Crazzolara01} have
noticed the same deficiency in strand spaces, and have derived a
similar notion of conflict between strands. 
}

Define an \emph{extended strand space} as a tuple
$(\Sigma,\cA,A,\xor)$ consisting of a strand space
$\Sigma$, a set $\cA$ of agents, an agent assignment $A$ from strands
to agents, and a set $\xor = \{\xor_a: a \in \cA\}$
of symmetric relations, indexed by
agents, such that $\xor_a\subseteq A^{-1}(a)\times A^{-1}(a)$. The
intuition is that if two strands $s_1$ and $s_2$ corresponding to the
same agent $a$ are such that $\xor_a(s_1,s_2)$,
then $s_1$ and $s_2$ 
\emph{conflict}; they
cannot both appear in the same bundle of
$(\Sigma,\cA,A,\xor)$. 
Formally, it is
sufficient to refine the definition of a bundle. If 
$(\Sigma,\cA,A,\xor)$
is an extended strand space, a \emph{bundle} $\cC$ of
$(\Sigma,\cA,A,\xor)$
is, as in the case of standard strand spaces, a
subgraph $(\cN_{\cC},(\strandsend_{\cC}\cup\strandnext_{\cC}))$ of the 
strand space $\Sigma$, satisfying 
B1--4 and, in addition:
\begin{enumerate}
\item[B5.] if 
$A(s_1)=A(s_2)=a$ and $\xor_a(s_1,s_2)$,
then it is not the case that both 
$\cheight{\cC}(s_1) 
\ge 1$ and 
$\cheight{\cC}(s_2)
 \ge 1$
(intuitively, if 
$\xor_a(s_1,s_2)$,
then $s_1$ and $s_2$ cannot both appear in $\cC$).
\end{enumerate}
We can similarly define an infinite bundle as a subgraph satisfying
B2--5; the notion of \joeheight{} remains unchanged.

Clearly, every bundle in an extended strand space
$(\Sigma,\cA,A,\xor)$
is a bundle of $\Sigma$, 
since properties B1--4 still hold.
Moreover, properties such
as the penetrator bounds proved 
by THG
carry over to extended strand spaces.

We
now consider translations from extended strand spaces to
strand systems and back. We first need to check that the construction
of Section~\ref{s:tomas} that translates a strand space into a strand
system applies to extended strand spaces.
Since a bundle in an extended
strand space is a bundle in the underlying strand space, we define the 
set $\chains(\Sigma,\cA,A,\xor)$ 
as the subset of $\chains(\Sigma,\cA,A)$ 
where each chain is taken over bundles in the extended strand space. As in
Section~\ref{s:tomas}, we can associate a run $r^C$ with every chain
of $\chains(\Sigma,\cA,A,\xor)$, and we define
$\R(\Sigma,\cA,A,\xor)=\{r^C:C\in\chains(\Sigma,\cA,A,\xor)\}$. The
analogue of Theorem~\ref{t:general} can be proved.

\thm\label{t:extgeneral} $\R(\Sigma,\cA,A,\xor)$ is a strand system. 
\ethm
Therefore, extended strand spaces can be translated
into strand systems in such a way that chains 
correspond 
to the runs of the system.  
We abuse notation and call this family of translations $T_A$ as well,
where
$A$ is an agent assignment (although now the domain of
$T_A$ is extended strand spaces
over the agent assignment $A$).
However, the maps $T_A$ are now onto,
and the following theorem holds.

\thm\label{t:xor} Given a strand system $\cR$ over $\cA$, there exists an 
extended
strand space 
$(\Sigma,\cA,A,\xor)$ such that 
$T_A(\Sigma,\cA,A,\xor)=\cR$.
\ethm
Extending the strand space model with a notion of per-agent conflict
relation is not the only way to extend the model to match the
expressiveness of strand systems. 
For instance, it is possible 
to introduce a more general form of conflict 
specifying
that an arbitrary pair of 
strands in a strand space cannot appear in any bundle. This notion of
conflict does not require the introduction of agents in the
strand-space framework. On the other hand, this extension is actually
more 
expressive than strand systems as defined in this paper. 
For example, it is possible to say that a particular history of agent 1
and another history of agent 2 do not occur in the same run, something
which cannot be done in a strand system.
While it is 
straightforward to augment strand systems to 
capture this stronger notion of conflict, it is not clear that such a
notion is of particular interest.
\commentout{
\thm 
Let $\cR$ be a strand system over a set $\cA$ of
agents. Define the strand space $\Sigma_\cR$ as:
\[\Sigma_\cR = \{s_{a,r,m} ~|~ a\in\cA, r\in\cR, m\geq 0\},\]
with a trace mapping $\mbox{tr}(s_{a,r,m})=r_a(m)$. 

Clearly, we get a great many similar strands (similar, in the sense
that they have the same trace). We now define the conflict relation
$\xor$ on $\Sigma_\cR$ to allow only bundles that ``span'' the strand
corresponding to a given run: $s_{r,a,m}\xor s_{r',a',m'}$ if and only 
if $r\not=r'$ or $m\not=m'$. In other words, the only strands that do
not conflict are strands of the form $s_{a,r,m}$ and $s_{a',r,m}$
corresponding to different agents in the same run $r$ up to the same
time $m$.  

It remains to show that $\Sigma_\cR$ translates back to $\cR$ with
agent set $\cA$ and agent assignment $A$ given by $A(s_{r,a,m}) = a$.

\pro
The system $\cR$ is generated by $V_a=\{\evtsseq_a(C|_n) ~|~
C\in\chains(\Sigma_{\cR}),n\geq 0\}$; that is, $\cR =
\cR(\seq{V_a}{a\in\cA})$. 
\epro
\prf
\eprf
}%

\section{From protocols to systems}\label{s:protocols}

Up until now, we have assumed that our strand spaces and systems were
simply given. This is the assumption typically made in the strand
spaces literature. In practice, however, strand spaces and systems
arise out of the agents executing protocols. In this 
section, we review the basics of how to derive a strand
system from an explicit protocol. This is a straightforward 
application of the techniques of \cite{Fagin95}. We then explore, using
this approach, why the strand spaces approach is successful when
dealing with typical protocols,
despite the restrictions pointed out in Section~\ref{s:frommas}. 
Roughly speaking, the strand systems
that are generated from typical protocols are images of strand spaces
via our translation. In other words, typical authentication protocols
do not lead to systems that cannot be expressed as strand spaces;
these protocols do not make choices. This result is not surprising
given our previous discussion in Section~\ref{s:frommas}, but it does
formally ground our intuition.
(On the other hand, we should point out that modern security protocols
often involve choosing subprotocols.) 

Intuitively, a protocol for agent $a\in\cA$ is a description of what
actions $a$ may take as a function of her local state. 
What actions are we to allow in our protocols? This question ties in
with the computational model implicitly assumed by strand
spaces. Notice that in strand spaces, the receiver of a message cannot
be specified.  Indeed, there is an edge $n_1\strandsend n_2$
between all nodes of the form $+u$ and $-u$.%
\footnote{We 
could
extend the notion of 
strand spaces to allow a specification of which
$\strandsend$ edges should be included; we could also add a
``tagging'' mechanism to messages.  However, our interest here is not in
extending the strand space formalism, but in modeling
strand spaces as defined by THG.} Therefore, we will consider a model
where a $\send$ action sends a message nondeterministically to any
agent. The only other action we allow beyond $\send$ actions is a ``do
nothing'' $\nop$ action. 
(We could also incorporate other actions, such as
choosing keys, or tossing coins to randomize protocols.) For
simplicity, we take actions to be deterministic, although protocols
themselves can be nondeterministic.
In other words, we will not consider an action such
as ``send some 
nondeterministically
chosen message $u$'', but rather a
protocol chooses 
nondeterministically
among the actions ``send $u_1$'', ``send $u_2$'', etc. 

We can formalize this intuition as follows. Fix a set $L_a$ of
local states for agent $a$ (the local states that arise in some
system) and a set $\act_a$ of possible actions that agent $a$ can take. A 
\emph{protocol} $P_a$ for agent $a$ is a function that associates with every
local state in $L_a$ a nonempty subset of actions in
$\act_a$. Intuitively, $P_a(\sigma)$ is the set of actions that agent $a$
may perform in local state $\sigma$. Notice that $a$'s actions can 
depend only on her local state. If $P_a$ prescribes a unique action for $a$ 
at each local state, then $P_a$ is said to be \emph{deterministic}. 
To
consider the effect of all the agents' protocols on the global state
of the system, 
define a \emph{joint protocol} $\<P_a : a\in\cA\>$ to be a tuple 
consisting of a protocol for each of the agents.  A joint protocol maps
a global state to a set of \emph{joint actions}, where a joint action is
a tuple consisting of an action in $\act_a$ for each agent $a$.  Define
$\<P_a: a \in \cA\>(\<\sigma_a: a \in \cA\>) = \{\sact_a: a \in \cA,\,
\sact_a \in P_a(\sigma_a)\}$.

Joint actions transform global states. 
Their effect is captured by a transition function $\tau$ mapping
global states to sets of global states.\footnote{In \cite{Fagin95},
the transition function is taken to be a function from
global states to global states. 
The nondeterminism inherent in our definition is avoided 
by taking
an \emph{environment}
as an extra agent in systems. 
For simplicity, we have not considered environments in this
paper.}
Given a joint protocol, a transition function, and
a set of initial global states, we can generate a system in a
straightforward way. Intuitively, the system consists of all the runs
that are obtained by running the joint protocol from one of the
initial global states. More formally, say that run $r$ is
\emph{consistent with joint protocol $P$ given transition function $\tau$}
if it could have been generated by $P$, that is, for all $m$, $r(m+1)$
is the result of applying a joint action $\sact$ that could have been
performed according to protocol $P$ to $r(m)$. (More precisely, there
exists a joint action 
$\sact=\<\sact_a : a\in\cA\>$
such that  $\sact_a\in P_a(r_a(m))$ and $r(m+1)\in\tau(\sact)(r(m))$.)
For a joint protocol $P$, a transition function $\tau$, and a set of
initial state $I$, let $\R(P,\tau,I)$ be the set of all runs $r$
consistent with $P$ given $\tau$ such that $r(0)\in I$.  

We saw in Section~\ref{s:frameworks} that strand systems are
asynchronous systems; they do not provide any guarantee either with
respect to the time it takes to deliver a message, or with respect to
the relative rates at which agents perform actions.  
We can capture this asynchrony 
by 
using an appropriate transition function.  First, note
that the only action in strand systems is that of
sending a message. Also, 
the local states of an agent is its history.
When agent $a$'s component of a joint action $\sact$ is a $\send(u)$
for some message $u$, 
$\tau(\sact)$ may or may not actually send the message.
If the message is not sent, the agent's local state is
unchanged, and hence the agent's protocol will allow the message 
to be re-sent. 
A message can be delivered in any round after it is sent (and may never
be delivered at all).
To capture this, we use
the \emph{strand transition function} $\tau_P$ of a joint protocol $P$,
defined 
as follows: a global state $\<\sigma'_a : a\in\cA\>\in
\tau_P(\<\sact_a : a\in\cA\>)(\<\sigma_a : a\in\cA\>)$ iff
for all $a\in\cA$, $\sigma'_a$ is either $\sigma_a$,
$\sigma_a\cdot\sent(u)$  
(only if $\sact_a=\send(u)$), 
or $\sigma_a\cdot\receive(u)$ (only if 
$\sent(u)\in \sigma'_b$ for some $b$).
We can check that using the strand transition function does indeed
yield a strand system. Note that in a strand system, each agent is
assumed to start with an empty initial local state. Hence, we always
take the set of initial global state to be $I_0$, which contains only
the empty global state. 
\thm\label{t:strandprotocol}
$\R(P,\tau_P,I_0)$ is a strand system.
\ethm

We now have all the machinery to explain why strand spaces can be 
suitable for modeling typical protocols found in the
literature. (We say \emph{can} because part of the suitability issue
depends on the actual properties we want to prove, as we will discuss
in Section~\ref{s:discussion}.) What do we mean by ``suitable for
modeling''?  We have described above how protocols generate systems in a 
natural way. Furthermore, we saw in Section~\ref{s:frommas} that
not every strand system arises as a translated strand space. In those
cases, the most natural strand space 
allows bundles
that do not correspond to 
states that actually occur  in the system. A strand space is suitable as a model
for a system if the translation of that strand space (under some agent 
assignment) yields the system. We interpreted our results of
Section~\ref{s:frommas} as showing that strand spaces could not
express choice
(and other related forms of nondeterminism).
Hence, intuition would indicate that if a protocol
avoids such nondeterminism, strand spaces should be suitable for modeling
the generated system.

It turns out that capturing this intuitive notion is not so easy.
The restrictions that must be imposed to ensure that the
generated strand system can be expressed as a strand space are
nontrivial. 
To make them precise, we need a few definitions.
For 
a class $\cP$ of protocols, a joint protocol $\<P_a : a\in\cA\>$ is
\emph{decomposable} into protocols in $\cP$ if for each agent $a$, we
can find protocols $P_a^1, P_a^2, \ldots$ in $\cP$ such that for all
global states $\<\sigma_a : a\in\cA\>$ we have $P_a(\sigma_a)=\cup_i
P_a^i(\sigma_a)$.
In other words, a decomposable joint protocol can be understood as
each agent running a set of protocols in a given class $\cP$.  A
deterministic protocol $P$ is \emph{monotone} if there exists 
events $e_1,e_2, \ldots$ (the sequence may be finite or infinite), 
such that  
for any local state $\sigma$, we have
$P(\sigma)=\{\send(u)\}$, 
if $e_{i+1} = \sent(u)$ and $i$ is the largest index such that 
$e_1,\ldots,e_i\in\sigma$; and $P(\sigma) = \{\nop\}$
otherwise. 
Informally, a deterministic protocol is monotone if the possible
action at a state depends only on whether or not a given set of events
has occurred. (Other events in the state does not affect the possible
action.) For example, a monotone protocol may be of the form: send
message $u_1$, wait for message $u_2$, send message $u_3$. This is
monotone in our sense, since $u_1$ is sent not matter what, and
$u_3$ is sent if and only if $u_2$ is received.
This means that if 
certain messages are sent by $a$ in one run of the
protocol, they must be sent by $a$ in all runs of the protocol. 
This must be true even if the protocol is run in parallel with other
protocols (or other instantiations of the same protocol).
In a 
sense, monotone protocols don't ``interact''; if an agent is running
multiple monotone protocols at the same time, they cannot keep each
other from proceeding.

It is not hard to show that neither of the two systems that we showed in 
Section~\ref{s:frommas} were not representable as strand spaces are
generated by monotone protocols.  The first protocol (where there was a
nondeterministic choice) is not itself monotone, since
agent 2 nondeterministically chooses to send a message
to agent 1 or agent 3.  Nor can it be split into two deterministic
monotone protocols, 
one to communicate with agent 1 and one to communicate with agent 3.
Sending a message in one protocol would prevent a message being
sent in another, and thus neither of the two deterministic protocols is
monotone.  Intuitively, the two protocols
interact, something disallowed by monotonicity. 
Agent 2's protocol in the $\mathit{nack}/\mathit{ack}$ example is
not monotone either.  The obvious sequence of events for agent 2's
protocol is $\sent(\mathit{nack}), \receive(u), \sent(\mathit{ack})$,
but this does capture the protocol, since in runs where 2 receives $u$
before sending the $\mathit{nack}$ message will not arise in the
protocol corresponding to this sequence.

The following theorem shows that joint
protocols decomposable into monotone protocols can indeed be modeled
by strand spaces: 

\thm\label{t:monotone}
If $P$ is a joint protocol decomposable into monotone
protocols, then there exists a strand space $\Sigma$ and an agent
assignment $A$ such that $T_A(\Sigma)=\R(P,\tau_P,I_0)$.   
\ethm

Theorem~\ref{t:monotone}
somewhat explains why the restrictions on the modeling
power of strand spaces we pointed out in Section~\ref{s:frommas} are
not an issue when analyzing security protocols of the kind typically
found in the literature.  These protocols are monotone. 
Note that the penetrator in strand spaces analyses also
runs a protocol that is a union of monotone protocols: send a new
message, send a concatenation of two received messages, send part of a
received message. These monotone protocols correspond exactly to the
so-called penetrator strands.

\section{Discussion}
\label{s:discussion}

In this paper, we have investigated
the relationship
between strand spaces and multi-agent systems. 
Our results show that strand spaces are strictly less expressive than 
strand systems, a subclass of multi-agent systems that seems
to capture the assumptions underlying strand 
spaces,
in two quite distinct respects.
The first is that strand spaces cannot express choice, the fact that
exactly one of two possible behaviors is chosen.  The second is that 
strand spaces have no notion of agents.

How serious are these
two issues?  That depends, of course, on what we are trying to prove.
Consider first the inability of strand spaces to express choice.
In \cite{Thayer99}, the types of 
properties proved 
typically have the form ``for
all bundles in the strand space, X happens''. One way to
interpret our result of Section 
\ref{s:frommas} is that when a strand space is used to model
a system, some of the bundles may not correspond to situations that 
actually arise in the system---those bundles can be seen as
``impossible'' bundles. 
This is not a problem, of course, if the property of interest in fact
holds in the larger system.  However, this may not always be the case.
For example, we may well want to prove that a property like ``agent 2
sends at most one message'' holds in all executions of a protocol.
If the protocol also has the property that agent 2 can send messages to
either 1 or 3 (as is the case in the protocol described by the system
$\cR_1$ in Section~\ref{s:frommas}), then the fact that agent 2 sends at
most one message in every execution of the protocol will simply not be
provable in the 
strand-space framework.
On the other hand, as we saw in Section~\ref{s:protocols}, if we
consider only strand systems generated from protocols decomposable
into monotone protocols, a fairly  restrictive class of protocols,
then we know that there is a strand space modeling the system that
does not contain any such ``impossible'' bundles. 

The runs of a strand system can be viewed as a linearization of bundles,
that is, an explicit ordering of the actions performed by agents in
different bundles.   
THG
suggest that results about strands can be imported to runs.  For
example, 
on page 226,
they say   ``[Alternatively,] results about authentication 
protocols proved in a strand space context can be imported into the
more usual linear models  by linearizing the bundles.''
Our results point to subtleties in doing this.
More precisely, while results about strands can be imported to results
about runs (the runs that arise from translating the strand space to a
system), the converse may not be true, depending on the expressiveness
of the language.

Turning to the issue of agents, 
the  strand-space framework 
assumes that messages relayed between strands form the only means of
exchanging information between strands. In other words, there is no
shared state between strands. Therefore, for all intents and purposes, 
we can imagine that every strand is executed by a different agent. 
On the other hand, if the same agent is
executing two strands then, intuitively, it should know whatever is
happening on both strands, without requiring communication between 
them.
Furthermore,
as soon as one wants
to analyze the properties of strand spaces using 
belief and knowledge,
agents
to which 
the
knowledge can be ascribed
are needed.
But even without bringing in knowledge, 
we need to be careful in interpreting security results proved under the
assumption that different agents perform different strands.
Clearly this assumption is not, in general, true.
Ideally, security protocols should be proved
correct under any ``reasonable'' assignment of agents to roles
in the security protocol. 
At the very least it should be clear under which assignments the result
holds.
For instance, it is known that methods for the analysis of
cryptographic protocols that fail to handle multiple roles for the
same agent do not yield dependable results, as they may not reveal
\emph{multi-role flaws}. Snekkenes \citeyear{Snekkenes92} studies such
flaws in the context of various cryptographic protocol logics.
Multi-role flaws commonly arise when a cryptographic protocol logic
implicitly assumes that if 
an agent $a$ takes on a role $A$ in some session, then he will not also
take on another role $B$ in some different session.
This assumption is often a consequence of the identification of the notions
of role and agent. Snekkenes shows that reasonable protocols 
that can
be proved correct under
the assumption that an agent takes on the same role in all sessions are
flawed if this assumption is dropped.
Recent work on analyzing mixed protocols using strand spaces
\cite{Thayer99a} shows that strand spaces can be extended to
deal with what essentially amount to multi-role flaws. However, the
approach often requires \emph{phantom messages} (messages that are not 
actually exchanged during runs of the protocols) to carry state
information between the different protocol strands corresponding to
the same agent.  

Some of the topics we have explored in this paper appear in various
forms in other work. For example, Cervesato \emph{et al.}
\citeyear{Cervesato00} define a notion of \emph{parametric strand},
essentially a strand where messages may contain
variables. Parameterized strands correspond to roles, which are
implicit in the original work on strand spaces. 
The work of Cervesato \emph{et al.} also deals with
the evolution of the system described by a strand space; they define a 
one-step transition between bundles. The transition is reminiscent of
the one we describe in Section~\ref{s:tomas}, but is restricted to
extending a single strand at a time. (They also allow actions specific to
their formalization, such as the instantiation of a strand from a
parametric strand.)

The set of runs in the system and the agent assignment
are particularly significant 
when 
we consider specifications that are not {\em
run-based} \cite{Fagin95,Halpern00}.
A run-based specification
is checked on a per-run basis. For example, 
``agent 2 sends at most 1 message'' is a run-based specification:
given a run, one can check 
whether the property holds for that
run.
A run-based specification holds
for a set of runs if it holds for all runs in the set. In contrast, a
\emph{knowledge-based specification} \cite{Fagin95,Halpern00}
such as ``after running the protocol, agent 2 knows $X$'' 
cannot be checked on a per-run basis, as 
it relies on the set of runs \emph{as a whole} to verify the property.
It holds if, in all runs in the system that agent 2
considers possible after running the protocol,
$X$ holds. Clearly it does not suffice to look at an individual run to
determine whether such a property holds.
Similarly,
probabilistic specifications like ``$X$ holds in at most 3\% of
the runs'' also depend on the whole system and cannot be checked simply by
examining
individual runs.

Typical specifications in the security literature are
safety properties (in the sense of Alpern and Schneider
\citeyear{AlSch85}, ``bad things don't happen''), and hence are
run-based. Run-based specifications have the property that if
they hold in a system, they hold in any subset of the runs of the
system. 
It is ``safe'' to prove that a run-based specification holds of a strand
space which translates to a superset of the intended system.  Proving
that the property holds for ``impossible'' runs does not hurt.
This is not the case for properties that are not run-based.
We believe that knowledge-based specifications,
as well as probabilistic ones, will play a significant role in the
design and analysis of security protocols. 
Fairness is a good example. 
A protocol
is \emph{fair} if intuitively no protocol participant can gain an
advantage over other participants by misbehaving. In the context of
fair exchange protocols \cite{Asokan98,BenOr90,Shmatikov00}, 
where
two agents exchange one item for another, fairness ensures that
either each agent receives the item it expects, or neither receives
any information about the other's item. This notion of ``not receiving 
any information'' can be interpreted as meaning that no knowledge is
gained.
Our results
suggest that strand spaces, as currently defined, will have difficulty
handling 
such specifications.
We should point out that it is straightforward to reason about
knowledge in the context of strand spaces. For instance, Syverson
\citeyear{Syverson99} describes a framework where the set of bundles in a
strand space is viewed as a set of possible worlds.  He associates
with every strand in the strand space a principal, as we do, and uses
this setting to provide a model for the knowledge of principals. As
his framework is directly based on strand spaces, it suffers from the
same expressiveness problems we pointed out in
Section~\ref{s:frommas}. 
This emphasizes that the problem we point
out is not a problem of how to express knowledge in strand
spaces. Rather, it is purely a problem with expressiveness of the
\emph{models} allowed in the strand-space framework. 

Despite these criticisms, we feel strand spaces are an important and
useful formalism.  They can be used to provide simple, transparent proofs of
run-based properties.  Our results suggest it is worth exploring their
limitations and the extent to which extensions of strand spaces (such as
the extended strand spaces 
introduced here) retain these
properties.  

\appendix

\newcommand{\om}{\{\!\{}
\newcommand{\cm}{\}\!\}}

\section{Proofs}

\othm{t:general} $\R(\Sigma,\cA,A)$ is a strand system. 
\eothm

\prf Let $V_a$ consist of all the histories $r_a(m)$ for $r \in
\R(\Sigma,\cA,A)$.  Let $\R'$ be the strand system generated by 
the sequence $\<V_a : a\in\cA\>$. 
To show that $\R(\Sigma,\cA,A)$ is a strand system, it clearly suffices
to show that $\R(\Sigma,\cA,A) = \R'$. 
It is easy to check from the construction that every
run in $\R(\Sigma,\cA,A)$ satisfies MP1--3, and thus is in $\R'$. 
This shows that  $\R(\Sigma,\cA,A) \subseteq\R'$. 

To show that $\R'\subseteq\R(\Sigma,\cA,A)$, let $r$ be a run in
$\R'$. 
We know that $r$ satisfies MP1--3, and that $r_a(m)\in V_a$ for
all $m\geq 0$. We need to construct a chain $C$ such that
$r_a(m)=\Hist^m_a(C)$
for all $a \in \cA$.
Unfortunately, we cannot simply construct the
chain inductively,
bundle by bundle.
While this would work if different strands 
were associated with
different agents, in general, 
making the correct choice 
of strands at each step
(correct in the sense
that the construction will not get stuck at a later point) 
turns out to require
arbitrary lookahead into the run. 
Roughly speaking, this is because it is not clear which combination of
strands 
for agent $a$
to choose to make up 
$a$'s local state in a particular bundle.

Instead, we proceed as follows. Intuitively, we want to
determine for each agent which strand prefix to extend at
every step of the chain. Once we have found for each agent 
an appropriate way of extending strand prefixes at every step, it is not
hard to construct the bundles in the chain.

We start with some definitions. 
Given a node $\<s,k\>$ in $\Sigma$, let $\tr(s,k)$ be the prefix of
$\tr(s)$ of length $k$.  Given a bundle $B$ and an agent $a$, let
$$\Tr_a(B) =
\om \tr(s,k): 
\<s,k\> \in \cN_B, \, \<s,k+1\> \notin \cN_B,
\, k \ge 1,\, A(s) = a\cm,$$  
where we use the $\om\cm$ notation to denote multisets.
Thus, $\Tr_a(B)$ is the multiset consisting of all  the maximal prefixes
of strands associated with $a$ having at least one node in $B$.
Note that $\Tr_a(B)$ is a multiset, not a set. 
It is quite possible that 
there are distinct nodes $\<s,k\>$ and $\<s',k\>$ in $\cN_B$ such that 
$\tr(s,k) = \tr(s',k)$ and $\<s,k+1\>, \<s',k+1\> \notin B$.  In this
case,  $\tr(s,k)$ is listed at least twice in the multiset.
Given a multiset $M$ of sequences, let 
$\cB_a(M) = \{B: \Tr_a(B) = M\}$.
That is, $\cB_a(M)$ consists of all bundles where the actions performed
are precisely those specified by the sequences in $M$.

For each agent $a$, 
we 
inductively
construct the following tree, whose vertices are labeled by
multisets of sequences.
The root is labeled by the empty multiset.  
Suppose a vertex $u$ at level $m$ (that is, at distance $m$ from the root) is
labeled with the multiset $M$.  If $r_a(m+1)=r_a(m)$, then $u$ has a
unique successor, also labeled with $M$.  
If, on the other hand, $r_a(m+1)=r_a(m)\cdot e$ for some event $e$, then
let $t$ be the term corresponding to $e$ (i.e., if $e$ is $\sent(u)$
then $t$ is $+u$, and if $e$ is $\receive(u)$ then $t$ is $-u$).
For each sequence $S$ in $M$, let $M_S$ be the multiset that results
from replacing $S$ in $M$ by 
$S\cdot t$.
We construct a successor of
$M$ labeled $M_S$ if $\cB_a(M_S) \ne \emptyset$.  (If $\cB_a(M_S) \ne
\emptyset$ and there are several
occurrences of $S$ in $M$, then we construct one successor for each
occurrence.)  In addition, if $\cB_a(M \cup \om \<t\> \cm) \ne
\emptyset$, we construct a successor of $u$ labeled $M \cup \om \<t\>
\cm$.
Note that, for all multisets labeling a level-$m$ vertex, the set of
events specified by the sequences in $M$ are precisely those performed
in $r_a(m)$.

Our goal is to find an infinite path in this tree.  That such a path
exists is immediate from K\"onig's Lemma, once we show that the tree has
an infinitely many vertices, each with finite outdegree.

An easy induction shows that a multiset at level $m$ has at most $m$
elements (counted with multiplicity).   Moreover, it is immediate from
the construction that the outdegree of a 
vertex on the tree 
is at most one more than
the cardinality of the multiset labeling it.  Thus, it follows that the
outdegree of each vertex is finite.

Showing that the tree has an infinite number of 
vertices is also
relatively straightforward. 
We show by induction on $m$ that for all times $m$, 
if $r_a(m) = \Hist^{k}_a(C)$ and $C = B_0\mapsto B_1 \mapsto\ldots$,
then there is a vertex at level $m$ in the tree labeled by the multiset
$\Tr_a(B_k)$.  The base case is immediate, since $\Tr_a(\emptyset) = \om
\cm$ is the label of the root of the tree.  Now suppose 
that
the result holds
for $m$; we prove it for $m+1$.  Suppose that $r_a(m+1) =
\Hist^{k}_a(C)$.  Then there must be some $k' \le k$ such that 
$r_a(m) = \Hist^{k'}_a(C)$.  Moreover, either $\Hist^{k'}_a(C) =
\Hist^{k}_a(C)$, in which case $r_a(m) = r_a(m+1)$, or $\Hist^{k}_a(C)$ is
the result of appending one event, say $e$, to $\Hist^{k'}_a(C)$ and
$r_a(m+1)$ is the result of appending $e$ to $r_a(m)$.  If $C =
B_0\mapsto B_1 \mapsto \ldots$ then, by the induction hypothesis, there
is a vertex $u$ of the tree at level $m$ labeled by $M = \Tr_a(B_{k'})$.  If 
$r_a(m) = r_a(m+1)$, then $M = \Tr_a(B_k)$ is also the label of a
successor of $u$.  Otherwise, if $M' = \Tr_a(B_k)$, it is clear that
$M'$ is the result of extending one of the strands in $M$ by one node
(corresponding to event $e$).  Thus, $M'$ is the label of some successor
of $u$.  This completes the inductive step.  
Since $r$ is in $\cR'$, it follows that, for all $m$, there exists some
chain $C$ and $k$ such that $r_a(m) = \Hist^k_a(C)$.  Thus, there are
infinitely many vertices in the tree.

It now follows from K\"onig's
Lemma that there is an infinite path in the tree.
Thus, it follows that, for every agent $a$, there exists an infinite
sequence 
$M^a_0,M^a_1,\ldots$ of multisets, 
such that $\cB_a(M^a_k)\ne\emptyset$ for all $k$. 
We now construct a chain $C = B_0 \mapsto B_1 \mapsto \ldots$,
by building the bundle $B_k$ from the traces in $\{M^a_k : a\in\cA\}$. 
For each $a$  
and $k$, there is a bundle $B^a_k$ such that $\Tr_a(B^a_k) = M^a_k$.
Let $B_k$ consists of the nodes in $\cup_{a \in \cA} B^a_k$
(so that the strands associated with $a$ in $B_k$ are precisely
those associated with $a$ in $B^a_k$), adding
$\strandsend$ edges between 
corresponding nodes
according to MP2 in the run $r$. 
That $B_k$
is a bundle follows from the fact 
that 
every node appearing in a multiset
$M^a_k$ corresponds to an event in $r_a(k)$, by construction.
It should be clear that for all $k$, $B_k\mapsto B_{k+1}$, 
since
for each agent, the traces are extended by a single node, and we can pick the
bijection $f$ to map strands from $B_k$ to $B_{k+1}$ so that the
corresponding sequences in $M^a_k$ and $M^a_{k+1}$ match. 

A straightforward induction argument shows that the chain
$C=B_0\mapsto B_1 \mapsto \ldots$ is such that $r_a(m)=\Hist^m_a(C)$
for all $m\geq 0$.
\eprf

\othm{t:adequate}
Every global state of $\cR(\Sigma,\Sigma,\mathit{id})$ is message-equivalent
to a bundle of $\Sigma$ of finite \joeheight{}, and every bundle of
$\Sigma$ of finite \joeheight{} is message-equivalent to a global state of
$\cR(\Sigma,\Sigma,\mathit{id})$. 
\eothm

We first prove two lemmas about chains.  

\lem\label{l:chain-induces-width}
In a chain $C = B_0\mapsto B_1 \mapsto B_2\mapsto\ldots$, the
\joeheight{} of $B_n$ is at most $2n$.
\elem

\prf
We show this by induction on $n$. 
Clearly, the \joeheight{} of $B_0$ is $0$. Assume the result
holds for the bundle $B_m$. Consider the bundle $B_{m+1}$. 
\commentout{
Since
$B_m\mapsto B_{m+1}$, for every agent at most one strand (up to $f$
bijection) is extended, by either a $+u$ or a $-u$. Consider the set
$N=\{n_a:~a\in\cA'\}$ (with $\cA'\subseteq\cA$) of all the nodes added
from $B_m$ to $B_{m+1}$ (again, up to $f$ bijection). We show that the 
longest ``chain'' $n_1\leadsto n_2\leadsto\ldots$ among those nodes is
of length 2, and therefore any causal chain in $B_m$ can be extended
by at most 2. Consider any node $n\in N$. If $\term(n)=+u$, and there is an 
edge
from $+u\strandsend -u$ for some $-u$, the $-u$ must be taken from one 
of the nodes in $N$; otherwise, the $-u$ is already in $B_m$, and
therefore by B2 there is already a node $+u$ in $B_m$, and therefore
$n$ cannot also have an edge to $-u$ (again, by B2). So $-u$ must be a 
new node (in $N$), and moreover this $-u$ cannot must end the chain:
it cannot be followed by a $\Rightarrow$ node (since by assumption
strands are extended by at most one node), and it cannot be followed
by $\rightarrow$ node, since it is a receive. Thus, any causal chain
starting from $n$ has length 2.  Conversely, if
$\term(n)=-u$, there cannot be any node following it, so any causal
chain starting from $n$ has length 1. In any case, adding these causal
chain to any causal chain in $B_m$ yields a maximal causal chain
length of at most $2m+2 = 2(m+1)$, which  proves the induction step,
and proves the lemma.}
Since
$B_m\mapsto B_{m+1}$, there is a bijection $f$ such that 
$B_m \sqsubseteq_f B_{m+1}$.  Consider a causal path
$n_1\leadsto n_2\leadsto\ldots$ in $B_{m+1}$, where $\leadsto$ is either
$\strandsend$ or $\strandnext$.  We claim that it contains at most two
``new nodes'', that is, it
contains at most two nodes in $B_{m+1}$ not of the form $\<f(s),i\>$ for
some node $\<s,i\>$ in $B_m$; moreover, the ``new'' nodes must come
at the end of the causal path.  To see this, suppose that $n$ is a new
node on the path and $n \leadsto n'$ for some $n'$ on the path.
If $n'$ is not a new node, it cannot be the case that $n
\strandsend n'$ (for otherwise, by B2, $n$ would not be a new node), and
it cannot be the case that $n \strandnext n'$ (for otherwise, by B3,
$n$ would not be a new node).
Thus, $n'$ must be a new node.  It follows that all
the new nodes on the causal path must follow the old nodes on the path.
Now suppose that there are three new nodes on the path; then it must be
the case that there are three new nodes $n, n', n''$ such that $n
\leadsto n' \leadsto n''$.  It cannot be
the case that $n \strandnext n'$, for then $n$ and $n'$ are both on the
same strand, contradicting the assumption in the construction that at
most one new event is added per agent.  Similarly, it cannot be the case
that $n' \strandnext n''$.  Thus, we must have $n \strandsend n'
\strandsend n''$.  But then  $\term(n') = -u$ for some message $u$, and
it cannot be the case that $n' \strandsend n''$.  
Thus, it follows that the causal path has at most two new nodes. 
Since, by the induction hypothesis, there are at most $2m+1$ ``old''
nodes on the path, the path has 
at most $2m+3$ nodes and hence
length at most $2m+2$, as desired.
\eprf

Note that Lemma~\ref{l:chain-induces-width} does not depend on the 
assumption that each strand is associated with a distinct agent; the
following lemma does.

\lem\label{l:width-induces-chain}
If $B$ is bundle of finite \joeheight{}, then there exists bundles
$B_1,\ldots,B_k$ for some $k$ such that $B_0\mapsto B_1 \mapsto \ldots 
\mapsto B_k \mapsto B$. 
\elem

\prf First note that if $n$ is the last node on a
causal path in a bundle $B$ of maximum length, then either 
$\term(n)=-u$ for some $u$ or
$\term(n)=+u$ for some $u$ and there is no corresponding receive
node in $B$.

We now prove the result by induction on the \joeheight{} of $B$, that is
the length of the longest causal path. Clearly, if the \joeheight{}
of $B$ is $0$, then $B=B_0$. Otherwise, let $B'$ be the bundle derived 
from $B$ in the following way: for every strand $s\in\Sigma$, if the
last term of the prefix of $s$ in $B$ is $-u$ for some $u$
or if the last term is $+u$ and there is no corresponding $-u$ in $B$,
then let
$B'$ contain the prefix of $s$ 
that consists of every node in $s$ that is in $B$ but the last one;
otherwise, let $B'$ contain 
the same prefix of $s$ as $B$.
Clearly, $B'\mapsto B$. 
(Here we need the assumption that each strand is associated with a
different agent to ensure that in going from $B'$ to $B$, each agent
performs at most one action.)
Moreover, by the
initial observation, $B'$ does not include the last node of any causal
path of maximum length in $B$.
Therefore, the \joeheight{} of $B'$ is 
one less than the \joeheight{} of $B$.
Applying the induction hypothesis, we
get bundles $B_0\mapsto B_1\mapsto \ldots\mapsto B_k \mapsto B'
\mapsto B$, proving the result.
\eprf

\prf (Theorem~\ref{t:adequate})
If  $\<\sigma_s: s\in\Sigma\>$ is a global state in
$\cR(\Sigma,\Sigma,\mathit{id})$, then there must be some chain $C=B_0
\mapsto B_1 \mapsto \ldots$ and time $m$ such that $r^C(m) = \<\sigma_s
: s\in\Sigma\>$. 
By construction, $r^C_s(m)=\Hist^m_s(C)$, 
for each strand $s \in \Sigma$.  (Recall that $\cA = \Sigma$; we are
associating each strand with a different agent.)  Moreover,
$\Hist^m_s(C)$ is just the sequence of 
events performed in strand $s$ in $B_m$ (that is,  the prefix of 
$\tr(s)$ in $B_m$, under the standard correspondence between terms and 
events).  Therefore, $\<\sigma_s:s\in\Sigma\>$
is message-equivalent to $B_m$. Moreover, by
Lemma~\ref{l:chain-induces-width}, $B_m$ has finite \joeheight{}. 

Conversely, given a bundle $B$ of finite \joeheight{}, 
by Lemma~\ref{l:width-induces-chain}, there must exist $m$ and bundles 
$B_0, \ldots, B_m$ such that 
$B_0\mapsto \ldots\mapsto B_m\mapsto B$.  Thus, 
$C=B_0\mapsto \ldots\mapsto B_m\mapsto B\mapsto B\mapsto
B\mapsto\ldots$ is a chain.
Let $r^C$ be the run in
$\R(\Sigma,\Sigma,\mathit{id})$ corresponding to $C$. By the same
argument as above, $r^C(m+1)$ is message-equivalent to $B$. 
\eprf

\othm{t:notinimage}
There is no agent assignment $A$ and 
$A$-history preserving translation $T$ from strand spaces to
strand systems such that the strand system $\cR_1$ is in the image of $T$.
\eothm

\prf
By way of contradiction, suppose that 
$\Sigma$ is a strand space, $A$ is an agent assignment,
$T$ is a
translation which is 
$A$-history preserving, and 
$T(\Sigma) = \R_1$.
Since $T$ is $A$-history preserving, the presence of $r_1$ ensures that
there is a bundle $B_1$ in $\Sigma$ such that associated
with agent $2$ in $B_1$ is 
either a strand with prefix $\<+u,-v\>$ or
strands with prefix $\<+u\>$ and $\<-v\>$, and associated
with agent $1$ in $B_1$ there is either a strand with prefix $\<-u,+v\>$ or
strands with prefix $\<-u\>$ and $\<+v\>$. Similarly, 
the presence of $r_2$ in $\R_1$ guarantees that there is a bundle $B_2$
in $\Sigma$ such that
associated with agent $2$ in $B_2$ is either a strand with prefix
$\<+x,-y\>$ or strands with prefix $\<+x\>$ and $\<-y\>$, 
and associated with agent $3$ is either a strand with prefix
$\<-x,+y\>$ or strands with prefix $\<-x\>$ and $\<+y\>$.
In all those cases, there must be a bundle containing nodes
with the terms $+u$, $-u$, $+v$, $-v$, $+x$, $-x$, $+y$, and $-y$. The nodes
$+u$, $-v$, $+x$, and $-y$ are all on strands associated with agent 
$2$.
Since $T$ is $A$-history preserving, there
must  be a run in $\cR_1$ that contains four events for agent
$2$. This is a contradiction. 
\eprf

\othm{t:extgeneral} $\R(\Sigma,\cA,A,\xor)$ is a strand system. 
\eothm

\prf The proof is similar to that of Theorem~\ref{t:general}. We
simply need to check that when we are proving the
$\R'\subseteq\R(\Sigma,\cA,A,\xor)$ inclusion and constructing 
each bundle $B_k$ in the chain $C$ from the collection of traces
$\{M^a_k : a\in\cA\}$, each bundle is in fact a bundle in the extended 
strand space sense. This follows from the fact that we can choose for
each agent $a$ the strands making up the bundle in such a way that
none of the strands conflict, since we assumed that
$\cB_a(M^a_k)\ne\emptyset$ for $M^a_k$, and therefore there must exist
strands with the appropriate prefixes that do not conflict. 
\eprf

\othm{t:xor} Given a strand system $\cR$ over $\cA$, there exists an extended
strand space $(\Sigma,\cA,A,\xor)$ such that $T_A(\Sigma,\cA,A,\xor) = \cR$. 
\eothm

\prf
Let $V_a$ be a set of histories for each agent $a$, such that $\R$ is
generated by the sequence $\<V_a : a\in\cA\>$. 
Without loss of
generality, assume that each  $V_a$ is minimal, in the sense that
every history in $V_a$ actually appears in some run of $\R$. Define
the strand space 
$\Sigma=\{s^h_a :  a\in\cA,h\in V_a\}$
with a trace mapping $\tr(s^{\<e_1, \ldots, e_k\>}_a) =
\<t_1,\ldots,t_k\>$,
where if $e_i$ is $\sent(u)$, then $t_i$ is $+u$, 
and if $e_i$ is $\receive(u)$, then $t_i$ is $-u$. 

We define the conflict relation $\xor\subseteq\Sigma\times\Sigma$ to
ensure that bundles  include only one strand per agent. We set
$\xor(s^h_a,s^{h'}_a)$ if and only if $h\ne h'$. Intuitively, since a
bundle in $(\Sigma,\cA,A,\xor)$ can include only one strand per agent,  
and since strands correspond to possible local states, bundles
correspond to global states of the system $\R$. 

We show that $T_A$ maps $(\Sigma,\cA,A,\xor)$ to $\cR$, via the agent
assignment $A(s^h_a) = a$. This is a direct consequence of
the proof of Theorem~\ref{t:extgeneral}. We know that
$\cR$ is generated by $\<V_a : a\in\cA\>$. We also know that $T_A
(\Sigma,\cA,A,\xor)$ is generated by $\<V'_a:a\in\cA\>$, where $V'_a =
\{\Hist^m_a(C) : C\in\chains(\Sigma,\cA,A,\xor),m\geq 0\}$. Therefore, 
to show that $T_A(\Sigma,\cA,A,\xor)=\cR$, it is sufficient to show
that $V_a=V'_a$ for all $a\in\cA$.

Fix an agent $a\in\cA$. We first show that $V_a\subseteq V'_a$. Let
$h$ be a history in $V_a$, and let $r\in\R$ and $m\geq 0$ be such that 
$r_a(m)=h$. For each $k\leq m$, define $B_k$ to be the bundle formed
by the strands $\{s^{r_a(k)}_a:a\in\cA\}$, with edges between nodes on 
different strands given by MP2. (That $B_k$ is a bundle follows from
the properties MP1--3 on $r$.) It is easy to see that $B_k\mapsto
B_{k+1}$ (for $k=0,\ldots,m-1$).  Let $C$ be the chain $B_0\mapsto
\ldots \mapsto B_m \mapsto B_m \mapsto B_m\mapsto\ldots$. Then
$\Hist^m_a(C)$ is just the set of events corresponding to strand
$s^{r_a(m)}_a$=$s^h_a$, which is just $h$. Therefore, $h\in
V'_a$. Showing that $V'_a\subseteq V_a$, is similar. Let
$h$ be a history in $V'_a$, so that there exists a chain $C$ with
$h=\Hist^m_a(C)$ for some $m\geq 0$. By construction, there exists a
run
$r^C\in\R$ such that $r^C_a(m)=\Hist^m_a(C)=h$. Thus, $h$ is a local
state of some run in $\R$, and $h\in V_a$. 
\eprf

\othm{t:strandprotocol}
$\R(P,\tau_P,I_0)$ is a strand system.
\eothm
\prf 
Let $\R$ be $\R(P,\tau_P,I_0)$, and let $V_a$ consists of all the
histories $r_a(m)$ for $r\in\R$. Let $\R'$ be the strand system
generated by $\<V_a : a \in \cA\>$. It is sufficient to show that for
all runs $r$, $r\in\R$ iff $r\in\R'$.

First, assume that $r\in\R$, that is, that $r$ is consistent with $P$
given $\tau_P$, and that $r(0)\in I_0$. By
construction, $r_a(m)\in V_a$ for all $a$ and all $m$, and hence $r$
satisfies MP1. By definition of $\tau_P$, if $\receive(u)\in
r_a(m)$, then $\sent(u)\in r_b(m)$ for some $b$, 
and hence $r$ satisfies MP2. Finally, 
$r_a(0)$ is the empty sequence because $r(0)\in I_0$, and by
definition of $\tau_P$, $r_a(m+1)$ is either $r_a(m)$ or the result of
appending one event to $r_a(m)$, and hence $r$ satisfies
MP3. Therefore, $r\in\R'$.

Second, assume that $r\in\R'$, that is, $r$ satisfies MP1--3. The fact
that $r(0)\in I_0$ is a consequence of MP3: $r_a(0)$ is the empty
sequence for all $a$. To show consistency with $P$ 
given $\tau_P$, we exhibit, for any $m$, a joint action
$\sact$ such that $r(m+1)\in \tau_P(\sact)(r(m))$. Let
$r(m)=\<\sigma_a : a\in\cA\>$, and $r(m+1)=\<\sigma'_a :
a\in\cA\>$. For any $a\in\cA$, if $\sigma'_a=\sigma_a$, let $\sact_a =
\nop$; if
$\sigma'_a=\sigma_a\cdot\sent(u)$, let $\sact_a = \send(u)$; if
$\sigma'_a=\sigma_a\cdot\receive(u)$, 
let $\sact_a = \nop$
(by MP1 and MP2, we know that there must exist a $\sent(u)$ in
$\sigma'_b$ for some $b$).
We can check that $\sact=\<\sact_a : a\in\cA\>$ has
the required property. Hence, $r$ is consistent with $P$ 
given $\tau_P$, and $r(0)\in I_0$, and therefore $r\in\R$.
\eprf

\othm{t:monotone}
If $P$ is a joint protocol decomposable into monotone
protocols, then there exists a strand space $\Sigma$ and an agent
assignment $A$ such that $T_A(\Sigma)=\R(P,\tau_P,I_0)$.   
\eothm
\prf 
By definition, for all agents $a\in\cA$, there exist monotone
protocols $P_a^1, P_a^2, \ldots$. For each such protocol $P_a^i$, we
can find events $e_{a,1}^i,e_{a,2}^i, \ldots$ as in the definition.
Let $|P_a^i|$ denote the length of this sequence (which could be $\infty$).

Construct the strand space 
$\Sigma=\{s_a^{i,n} : a\in\cA, i\geq 1, 1 \le n \le |P_a^i|\}
\cup\{s_a^u : a\in\cA, u\in M\}$.
The strand $s_a^{i,n}$ corresponds to a prefix of length $n$ of the
events in the sequence for $P_a^i$.  More precisely, its trace mapping is
a trace mapping $\tr(s_a^{i,n})=\<t_1,\ldots,t_n\>$, where for all $1\leq
j \leq n$, if $e_{a,j}^i=\sent(u)$, then $t_j$ is $+u$, and if
$e_{a,j}^i=\receive(u)$, then $t_j$ is $-u$. 
The strands of the form $s_a^u$ are simple strands corresponding
to receiving message $u$; there is 
one such strand for
each message $u\in M$, and for each agent $a\in\cA$. The trace mapping is
simply $\tr(s_a^u)=\<-u\>$. (These strands will be used to account
for unsolicited messages delivered to agent $a$.) The agent assignment
$A$ is simply defined by taking $A(s_a^{i,n})=a$ and $A(s_a^u)=a$, as
expected. 

Recall from Section~\ref{s:tomas} that $T_A(\Sigma,\cA,A)$ maps to 
the set of runs $\{r^C : C\in\chains(\Sigma,\cA,A)\}$. We show that
this set of runs is just $\R(P,\tau_P,I_0)$. 

First, let $C$ be a chain in $\chains(\Sigma,\cA,A)$. Recall that
$r^C$ is the run with $r^C(m)=\<\Hist_a^m(C) : a\in\cA\>$. To show
that $r^C$ is in $\R(P,\tau_P,I_0)$, it suffices to show that $r^C$ is
consistent with $P$ given $\tau_P$, and that $r^C(0)\in
I_0$. The latter is an immediate consequence of the fact that
$r_a^C(0)=\Hist_a^0(C)=\<\>$. The former requires showing that for all 
$m$, we can find a joint action $\sact=\<\sact_a : a\in\cA\>$ such
that $\sact_a\in P_a(r_a^C(m))$ and
$r^C(m+1)\in\tau_P(\sact)(r^C(m))$. Fix  $m$. For an agent
$a\in\cA$, if $\Hist_a^m(C)=\Hist_a^{m+1}(C)$, take
$\sact_a=\nop$. 
Otherwise, observe that by construction of the
strand space $\Sigma$, there exist $j_1,j_2,\ldots$ such that
$\cup_i E_{a,j_i}^i\subseteq\Hist_a^m(C)$, and
$\Hist_a^{m+1}(C)-\Hist_a^m(C)$ is either $e_{a,j_i+1}^i$ for some
$i$, in which case, we take $\sact_a=\send(u)$ if
$e_{a,j_i+1}^i=\sent(u)$, and $\sact_a=\nop$ otherwise; if
$\Hist_a^{m+1}(C)-\Hist_a^m(C)=\{\receive(u)\}$ for some $u$,  we
take $\sact_a=\nop$. Finally, take $\sact=\<\sact_a :
a\in\cA\>$. It is not hard to check that this joint action has the desired
property. For example, if
$\Hist_a^{m+1}(C)-\Hist_a^m(C)=\{\sent(u)\}$, then
$\sact_a=\send(u)$. If
$\Hist_a^{m+1}(C)-\Hist_a^m(C)=\{\receive(u)\}$, then
$\sact_a=\nop$, and by the bundle
properties, there must have been a corresponding send appearing in
$\Hist_b^{m+1}(C)$ for some other $b$. In both cases, 
using strand transition function $\tau_P$ gives us the right result. Hence
$r^C\in\R(P,\tau_P, I_0)$ for $C\in\chains(\Sigma,\cA,A)$. 

Second, consider a run $r\in\R(P,\tau_P,I_0)$. We need to construct a
chain $C\in\chains(\Sigma,\cA,A)$ such that $r=r^C$. However, for the
same reasons as in the proof of Theorem~\ref{t:general}, we cannot
simply define the chain inductively. We use the same construction
as in the proof of Theorem~\ref{t:general}, which tells us how to
construct a chain; essentially, the chain is constructed by
``picking'' the right strands from bundles. However, to apply the
construction, we need to verify a few facts. From
Theorem~\ref{t:strandprotocol}, we know that $\R(P,\tau_P,I_0)$ is a
strand system, generated by $\<V_a : a\in\cA\>$, where $V_a=\{r_a(m) :
a\in\cA, r\in\R(P,\tau_P,I_0)\}$. The construction in the proof of
Theorem~\ref{t:general} relied on the fact that to every history $h\in
V_a$, there was a bundle in $\Sigma$ for which every event associated
with a strand of agent $a$ corresponded to an event in $h$. The same
holds in our setting: observe that $h\in V_a$ iff there exist
$j_1,j_2,\ldots$ such that $\cup_i E_{a,j_i}^i\subseteq h$ and
$h-\cup_i E_{a,j_i}^i$ is made up exclusively of $\receive$ events, by
definition of $\tau_P$ and the fact that $P$ is decomposable into
monotone protocols. For any such history $h$ there is a bundle $B$
with the following strand prefixes corresponding to agent $a$: for
each $i$, if $j_i=0$, then no node of $s_a^{i,n}$ is in $B$ (for any
$n$), while if $j_i=k$, then the first $k$ nodes of $s_a^{i,n}$ (for
any $n\geq k$) are in 
$B$; the events in $h$ unaccounted for by these strand prefixes are
$\receive$ events, for which corresponding strands of the form
$s_a^u$ are in $B$. (Whatever other strands are in $B$ are
unimportant, so simply take the downward closure of the given strand
prefixes.) Therefore, we can apply the construction in the proof of
Theorem~\ref{t:general} to get a chain $C=B_0 \mapsto B_1 \mapsto
\ldots$ with the property that 
$r_a(m) = \Hist_a^m(C)$. Hence, $r=r^C$. In other words,  $r\in
T_A(\Sigma,\cA,A)$.  
\eprf

\begin{acks}
We would like to thank Andre Scedrov for pointing us to fair exchange
protocols as a likely source of knowledge-based specifications in
security protocols. Vicky Weissman 
and Kevin O'Neill 
read a draft of this paper and provided numerous helpful suggestions.
\end{acks}

\bibliographystyle{acmtrans}
\bibliography{riccardo,joe,z} 

\begin{received}
???
\end{received}

\end{document}